\documentclass[12pt]{article}
\usepackage{amsmath,amsfonts,graphicx}

\makeatletter
\@addtoreset{equation}{section}
\makeatother

\begin{document}
\title{Induced vacuum energy-momentum tensor\linebreak in the background of a  $d-2$ - brane
\linebreak in $d+1$ - dimensional space-time}
\author{Yurii A. Sitenko\thanks{E-mail:
yusitenko@bitp.kiev.ua}\\
\it \small Bogolyubov Institute for Theoretical Physics,
\it \small National Academy of Sciences of Ukraine,\\
\it \small 14-b Metrologichna str., Kyiv 03143, Ukraine\\
\phantom{11111111111}\\
Volodymyr M. Gorkavenko
\thanks{E-mail: gorka@univ.kiev.ua}\\
\it \small Department of Physics, Taras Shevchenko National University of Kyiv,\\
\it \small 6 Academician Glushkov ave., Kyiv 03680, Ukraine}
\date{}
\maketitle

\begin{abstract}
Charged scalar field is quantized in the background of a static $d-2$ - brane
which is a core of the magnetic flux lines  in flat $d+1$ - dimensional
space-time. We find that vector potential of the magnetic core induces the
energy-momentum tensor in the vacuum. The tensor components are periodic
functions of the brane flux and holomorphic functions of space dimension. The
dependence on the distance from the brane and on the coupling to the space-time
curvature scalar is comprehensively analysed.
\end{abstract}
\renewcommand{\theequation}{\arabic{section}.\arabic{equation}}


\section{Introduction}
Since Casimir's seminal paper \cite{Cas} it has become clear that the effect of
external boundary conditions in quantum field theory can be exposed as the
emergence of a non-zero vacuum expectation value of the energy-momentum tensor
(see, e.g. Refs. \cite{Bir, Most}). This may have far reaching consequences; in
particular, the vacuum energy-momentum tensor serves as a source of
gravitation, and the so-called self-consistent cosmological models of the
Universe are proposed, where matter is absent and its role is played by the
vacuum quantum effects \cite{Star}.

In this respect it seems to be of interest to look for various situations when
the vacuum energy-momentum tensor is calculable and finite. Let $X$ be the base
space manifold of dimension $d$ and $Y$ be a submanifold of dimension less than
$d$. The matter field is quantized under certain boundary condition imposed at
$Y$. In most implications of the Casimir effect, $Y$ is chosen to be noncompact
disconnected (e.g. two parallel infinite plates, as generically in Ref.
\cite{Cas}) or closed compact (e.g. box or sphere), see Ref. \cite{Most}. In
the present paper we choose $Y$ to be noncompact connected and possessing
dimension $d-2$, i.e. a $d-2$ - brane in $d$-dimensional space. If such a brane
is filled with the magnetic flux lines, then it can be regarded as a
generalization of the Bohm-Aharonov \cite{Aha} singular magnetic vortex in
$3$-dimensional space. If the matter field vanishes at $Y$, then the region
where the matter field is nonvanishing  (out of $Y$) does not overlap with the
region where the background magnetic field is nonvanishing (inside $Y$). Thus
there is no effect of the background field on the matter field in the framework
of classical theory, and such an effect, if exists, is of purely quantum
nature. Conventional Bohm-Aharonov effect pertains to the quantum-mechanical
framework \cite{Aha}. As is clear from the above, our interest will be in the
quantum-field-theoretical framework (i.e. vacuum polarization in the background
of the brane), which, therefore, may be generally denoted as the
Casimir-Bohm-Aharonov effect, see also Ref. \cite{Sit}.

Throughout the present paper, we restrict ourselves to the case of scalar
matter. A peculiarity of this case is that the energy-momentum tensor depends
on the coupling ($\xi$) of the scalar field to the scalar curvature of
space-time even then when space-time is flat. If scalar field is massless, then
conformal invariance of the theory is achieved at $\xi=\xi_c$, where \cite{Pen,
Cher, Cal}
\begin{equation}\label{intr1}
\xi_c=\frac{d-1}{4d}\,;
\end{equation}
note that $\xi_c$ varies from $0$ to $1/4$ when $d$ varies from
$1$ to $\infty$. Our analysis of vacuum polarization effects in
the background of the brane will be carried out for arbitrary
values of $\xi$; however results will be most impressive in the
case of conformal coupling, $\xi=\xi_c$. We shall find out
components of the induced vacuum energy-momentum tensor as
functions of the brane flux, distance from the brane and space
dimension. Expressions for components are especially simple in
form in the case of massless scalar field, whereas in the massive
case they are presented in terms of convergent integrals of the
Macdonald functions.

In the next section which is also introductory but more detailed,
a general definition of the energy-momentum tensor for the
quantized charged scalar field is reviewed and a starting
expression for its regularized vacuum expectation value in the
background of the brane is given. In Section 3 which is central
in the paper, the regularized vacuum tensor components are
computed, and in Section 4 the renormalized vacuum tensor
components are obtained. Various aspects of the latter result are
examined in the following sections: asymptotic behaviour at large
and small distances from the brane (Section 5), expressions at
fixed values of the brane flux when tensor components have
maximal absolute values (Section 6), dependence on the
$\xi$-parameter (Section 7). Finally, results are summarized in
Section 8. Some details in the derivation of results are outlined
in Appendices A and B.


\section{Energy-momentum tensor and its vacuum \\ expectation value}
The energy-momentum tensor for the quantized charged scalar field
$ \Psi(x) $ is given by expression
\begin{equation}\label{a1}
   T^{\mu\nu}=T^{\mu\nu}_{can}+\xi\left(g^{\mu\nu}\Box-\nabla^\mu\nabla^\nu-R^{\mu\nu}\right)
   \left[\Psi^\dag,\Psi\right]_+\,,
\end{equation}
where
\begin{equation}\label{a2}
   T^{\mu\nu}_{can}=\frac12\left[\nabla^\mu\Psi^\dag,\nabla^\nu\Psi\right]_++
 \frac12\left[\nabla^\nu\Psi^\dag,\nabla^\mu\Psi\right]_+-g^{\mu\nu}L\,,
\end{equation}
and
\begin{equation}\label{a3}
L=\frac12\left[\nabla^\mu\Psi^\dag,\nabla_\mu\Psi\right]_+-\frac12\left(m^2+\xi
 R\right)\left[\Psi^\dag,\Psi\right]_+\,,
\end{equation}
 $\nabla_\mu $ is the covariant derivative involving both affine and
 bundle connections,\linebreak\mbox{$\Box=\nabla_\mu\nabla^\mu $}
 is the covariant d'alembertian, $R^{\mu\nu}$ is the Ricci
 tensor and $ R=g_{\mu\nu}R^{\mu\nu} $ is the scalar curvature of
 space-time, signature of space-time metric $g_{\mu\nu}$ is chosen as $(+-\ldots-)$. Canonical tensor (\ref{a2}) is obtained by applying
 Noether's theorem to lagrangian (\ref{a3}), whereas tensor
 (\ref{a1}) is obtained by variating $L$ (\ref{a3}) over metric tensor
 $g_{\mu\nu}$.

 Eq.(\ref{a1}) can be rewritten in a form
\begin{equation}\label{a4}
   T^{\mu\nu}={\tilde T}^{\mu\nu}+\left(\xi-1/4\right)\left(g^{\mu\nu}\Box-\nabla^\mu\nabla^\nu\right)\left[\Psi^\dag,\Psi\right]_+
 -\xi R^{\mu\nu}\left[\Psi^\dag,\Psi\right]_+\,,
\end{equation}
where
\begin{equation}\label{a5}
   {\tilde T}^{\mu\nu}=T^{\mu\nu}_{can}+\frac14\left(g^{\mu\nu}\Box
   -\nabla^\mu\nabla^\nu\right)\left[\Psi^\dag,\Psi\right]_+
\end{equation}
 is the canonical tensor corresponding to lagrangian  $\tilde L$ which
 differs from $L$(\ref{a3}) by a total divergence:
\begin{equation}\label{a6}
  \tilde {L}=L-
 \frac14\Box\left[\Psi^\dag,\Psi\right]_+=-\frac14\left[\Psi^\dag,\Box\Psi\right]_+-\frac14\left[\Box\Psi^\dag,\Psi\right]_+-\frac12\left(m^2+\xi
 R\right)\left[\Psi^\dag,\Psi\right]_+\,.
\end{equation}
Both $L$(\ref{a3}) and $\tilde {L}$(\ref{a6}) yield the same
equations of motion,
\begin{equation}\label{a7}
 \left[\Box+\left(m^2+\xi R\right)\right]\Psi=0,\qquad\left[\Box+\left(m^2+\xi
 R\right)\right]\Psi^\dag=0\,,
\end{equation}
but  $\tilde {L}$ is strictly vanishing on the solutions to the equations of
motion. In fact, $\tilde {L}$ is used as a lagrangian in the path integral
approach to quantization, since namely $\tilde {L}$ is directly related to the
inverse propagator of the quantized scalar field.

If $\Psi$ is a solution to Eq.(\ref{a7}), then one gets
\begin{equation}\label{8}
 {\tilde
 T}^{\mu\nu}=\frac12\left[\nabla^\mu\Psi^\dag,\nabla^\nu\Psi\right]_++
\frac12\left[\nabla^\nu\Psi^\dag,\nabla^\mu\Psi\right]_+-\frac14\nabla^\mu\nabla^\nu\left[\Psi^\dag,\Psi\right]_+\,,
\end{equation}
and
\begin{equation}\label{a9}
  g_{\mu\nu}T^{\mu\nu}=\left(\xi
 d-\frac{d-1}4\right)\Box\left[\Psi^\dag,\Psi\right]_++m^2\left[\Psi^\dag,\Psi\right]_+\,,
\end{equation}
where $d$ is the dimension of space. Thus, there is a distinctive
value of parameter  $\xi$ ($\xi=\xi_c$, see Eq.(\ref{intr1}))
under which the trace of the energy-momentum tensor becomes
proportional to the mass squared,
\begin{equation}\label{a10}
  \left.g_{\mu\nu}T^{\mu\nu}\right|_{\xi=\xi_c}=m^2\left[\Psi^\dag,\Psi\right]_+,
\end{equation}
and the tracelessness of $T^{\mu\nu}$ and conformal invariance
are achieved in the massless limit ($m=0$).

In the case of a static background $(\nabla_0=\partial_0, \quad g_{00}=1)$, the
operator of the quantized charged scalar field is represented in the form
\begin{equation}\label{a11}
  \Psi(x^0,\textbf{x})=\sum\hspace{-1.4em}\int\limits_{\lambda}\frac1{\sqrt{2E_{\lambda}}}\left[e^{-iE_{\lambda}x^0}\psi_{\lambda}(\textbf{x})\,a_{\lambda}+
  e^{iE_{\lambda}x^0}\psi_{-\lambda}(\textbf{x})\,b^\dag_{\lambda}\right].
\end{equation}
  Here, $a^\dag_\lambda$ and $a_\lambda$ ($b^\dag_\lambda$ and $b_\lambda$)
  are the scalar particle (antiparticle) creation and annihilation
  operators satisfying commutation relation; $\lambda$ is the set
  of parameters (quantum numbers) specifying the state;
  $E_\lambda=E_{-\lambda}>0$ is the energy of the state; symbol
  $\sum\hspace{-1em}\int\limits_{\lambda}$ denotes summation over discrete and
  integration (with a certain measure) over continuous values of
  $\lambda$; wave functions $\psi_\lambda(\textbf{x})$ are the
  solutions to the stationary equation of motion,
\begin{equation}\label{a12}
 \left\{-\triangle + [ m^2+\xi R(\textbf{x})]\right\}  \psi_\lambda(\textbf{x})=E^2_\lambda\psi(\textbf{x}),
\end{equation}
$ \triangle={\mbox{\boldmath $\nabla$}}^2$ is the covariant
laplacian. For components of the vacuum expectation value of the
energy-momentum tensor,
\begin{equation}\label{a13}
   t^{\mu\nu}=\left\langle \mathrm{vac} \left |T^{\mu\nu}\right|\mathrm{vac}\right\rangle,
\end{equation}
one gets expressions
\begin{equation}\label{a14}
   t^{00}=\sum\hspace{-1.4em}\int\limits_{\lambda}E_\lambda\psi^*_\lambda(\textbf{x})\,\psi_\lambda(\textbf{x})-(\xi-1/4)\triangle
  \sum\hspace{-1.4em}\int\limits_{\lambda}E^{-1}_\lambda\psi^*_\lambda(\textbf{x})\,\psi_\lambda(\textbf{x}),
\end{equation}
\begin{multline}\label{a15}
  t^{jj}{\,}^{'}=\frac12\sum\hspace{-1.3em}\int\limits_{\lambda}E^{-1}_\lambda
  \left\{\left[\vphantom{\nabla^j{\,}^{'}}\nabla^j\psi_\lambda(\textbf{x})\right]^*\left[\nabla^j{\,}^{'}\psi_\lambda(\textbf{x})\right]
  +\left[\nabla^j{\,}^{'}\psi_\lambda(\textbf{x})\right]^*\left[\vphantom{\nabla^j{\,}^{'}}\nabla^j\psi_\lambda(\textbf{x})
  \right]\right\}+\\
  +\left\{\frac14 g^{jj}{\,}^{'}(\textbf{x})\triangle-\xi\left
  [ g^{jj}{\,}^{'}(\textbf{x})\triangle
  +\nabla^j\nabla^j{\,}^{'}+R^{jj}{\,}^{'}(\textbf{x})\right]
   \right\}
  \sum\hspace{-1.4em}\int\limits_{\lambda}E^{-1}_\lambda\psi_\lambda^*(\textbf{x})
  \,\psi_\lambda(\textbf{x}),\\
  j,j{\,}^{'}=\overline{1,d};
\end{multline}
note that the $t^{0j}$ components are vanishing, and relations
$R^{00}(\textbf{x})=0$ and
\linebreak$\partial_0\left[\Psi^+,\Psi\right]_+=0$ have been taken
into account.

However, relations (\ref{a14}) and (\ref{a15}) can be regarded as
purely formal and, strictly speaking, meaningless: they are
ill-defined, suffering from ultraviolet divergencies. The
well-defined quantities are obtained by inserting an inverse
energy in a sufficiently high power
\begin{equation}\label{a16}
   t^{00}_{reg}(s)=\sum\hspace{-1.4em}\int\limits_{\lambda}
   E_\lambda^{-2s}\psi^*_\lambda(\textbf{x})\,\psi_\lambda(\textbf{x})-(\xi-1/4)\triangle
  \sum\hspace{-1.4em}\int\limits_{\lambda}E^{-2(s+1)}_\lambda\psi^*_\lambda(\textbf{x})\,\psi_\lambda(\textbf{x}),
\end{equation}
\begin{multline}\label{a17}
  t^{{jj}{\,}^{'}}_{reg}(s)=\frac12\sum\hspace{-1.3em}\int\limits_{\lambda}E^{-2(s+1)}_\lambda
  \left\{\left[\vphantom{\nabla^j{\,}^{'}}\nabla^j\psi_\lambda(\textbf{x})\right]^*
  \left[\nabla^j{\,}^{'}\psi_\lambda(\textbf{x})\right]
  +\left[\nabla^j{\,}^{'}\psi_\lambda(\textbf{x})\right]^*
  \left[\vphantom{\nabla^j{\,}^{'}}\nabla^j\psi_\lambda(\textbf{x})
  \right]\right\}+\\
  +\left\{\frac14 g^{jj}{\,}^{'}(\textbf{x})\triangle-\xi\left
  [ g^{jj}{\,}^{'}(\textbf{x})\triangle
  +\nabla^j\nabla^j{\,}^{'}+R^{jj}{\,}^{'}(\textbf{x})\right]
   \right\}
  \sum\hspace{-1.4em}\int\limits_{\lambda}E^{-2(s+1)}_\lambda\psi_\lambda^*(\textbf{x})
  \,\psi_\lambda(\textbf{x}),\\
  j,j{\,}^{'}=\overline{1,d}.\quad
\end{multline}
Sums (integrals) are convergent in the case of Re\,$s >  d/2$ .
Thus, summation (integration) is performed in this case, and then
the result is analytically continued to the case of $s = -1/2$.
This way of dealing with ultraviolet divergencies is known as the
zeta function regularization procedure \cite{Sal, Dow, Haw}.

It is amazing, as is already mentioned in Introduction, that the
energy-momentum tensor and, consequently, its vacuum expectation
value remain to be dependent on parameter $\xi$ even in the case
of flat space-time ($R=0$). If scalar field is quantized in the
background of a static magnetic field in flat space-time, then
the covariant derivative is defined as
\begin{equation}\label{a18}
\begin{array}{l}
{\mbox{\boldmath $\nabla$}}\Psi= \left({\mbox{\boldmath
$\partial$}}-i \textbf{V}\right)\Psi\,,\quad {\mbox{\boldmath
$\nabla$}}\Psi^\dag = \left({\mbox{\boldmath $\partial$}}
+i\textbf{V}\right)\Psi^\dag\,,\\
{\mbox{\boldmath $\nabla$}}\left[\Psi^\dag,\Psi\right]_+=
{\mbox{\boldmath $\partial$}}\left[\Psi^\dag,\Psi\right]_+\,,
\end{array}
\end{equation}
and the magnetic field strength takes form
\begin{equation}\label{a19}
  B^{ j_1\cdots j_{d-2}}(\textbf{x})=-\varepsilon^{ j_1\cdots j_d}
  \partial_{j_{d-1}}V_{j_d}(\textbf{x})\,,
\end{equation}
where $\textbf{V} (\textbf{x})$ is the bundle connection (vector
potential of the magnetic field), and $\varepsilon^{ j_1\cdots
j_d}$ is the totally antisymmetric tensor, $\varepsilon^{12\cdots
d}=1$.

In the present paper we consider the bundle curvature (magnetic
field strength) to be nonvanishing in the $d-2$ - brane (i.e.
point in the $d = 2$ case, line in the $d = 3$ case, plane in the
$d = 4$ case, and $d-2$ - hypersurface in the $d > 4$ case).
Denoting the location of the $d-2$ - brane by $x^1=x^2=0$, one
gets
\begin{equation}\label{a20}
  B^{3\cdots d}(\textbf{x})=2\pi\Phi\delta(x^1)\delta(x^2)\,,
\end{equation}
where $\Phi$ is the total flux (in the units of $2\pi$) of the
bundle curvature; then the bundle connection can be chosen in the
form:
\begin{align}\label{a21}
 V^1(\textbf{x})=-\Phi\frac{x^2}{\left(x^1\right)^2+\left(x^2\right)^2}\,,
 \quad &
 V^2(\textbf{x})=\Phi\frac{x^1}{\left(x^1\right)^2+\left(x^2\right)^2}\,,\nonumber
 \\
 V^j(\textbf{x})=0\,,\quad & j=\overline{3,d}\,.
\end{align}
The complete set of regular solutions to Eq.(\ref{a12}) in background
(\ref{a20})-(\ref {a21}) is given by functions (see, e.g., Ref \cite{Sit})
\begin{equation}\label{a22}
  \psi_{kn\textbf{p}\,}(\textbf{x})=(2\pi)^{\frac {1-d}2} J_{|n-\Phi|}(kr)
  e^{in\varphi}e^{i\textbf{p}{\textbf{x}}_{d-2}}\,,
\end{equation}
where
\begin{equation}\label{a23}
   0<k<\infty\,,\quad n\in\mathbb Z\,,\quad
   -\infty<p^j<\infty\,,\quad j=\overline{3,d}\,,
\end{equation}
$J_\mu(u)$ is the Bessel function of order $\mu $\,,
$r=\sqrt{{\left(x^1\right)}^2+{\left(x^2\right)}^2}$,
$\varphi=\arctan(\frac{x^2}{x^1})$,
 \mbox{${\textbf{x}}_{d-2}=(0,0,x^3,\ldots x^d)$}, and $\mathbb Z$ is the
 set of integers. Since solutions (\ref{a22}) correspond to the
 continuous spectrum $(E_{kn\textbf{p}\,}=\sqrt{{\textbf{p}}^2 + k^2 +
 m^2}>m)$, they obey orthonormality condition
\begin{equation}\label{a24}
  \int d^d x\,\,\psi^*_{kn\textbf{p}\,}(\textbf{x})
  \psi_{k^{'}n^{'}{\textbf{p}\,}^{'}}(\textbf{x})=
  \frac1k\delta(k-k^{'})\delta_{nn^{'}}\delta(\textbf{p}-{\textbf{p}}^{'})\,.
\end{equation}
Taking all the above into account, we get following expressions
for the nonvanishing components of the regularized vaccuum
expectation value of the energy-momentum tensor:
\begin{equation}\label{a25}
  t^{00}_{reg}(s)={\tilde t}{\,}^{00}_{reg}(s)+
  \left(\frac14 - \xi\right)\triangle_r{\tilde t}{\,}^{00}_{reg}(s+1)\,,
\end{equation}
\begin{multline}\label{a26}
 t^{rr}_{reg}(s)=(2\pi)^{1-d}\int d^{d-2}p\,\int\limits_0^\infty
 dk\,k\left({\textbf{p}}{\,}^2+k^2+m^2\right)^{-s-1}\sum_{n\in\mathbb Z}
 \left[\partial_rJ_{|n-\Phi|}(kr)\right]^2-\\
 -\left[\frac14\triangle_r-\frac{\xi}{r}\partial_r\right]
{\tilde t}{\,}^{00}_{reg}(s+1)\,,\,\,
\end{multline}
\begin{multline}\label{a27}
 t^{\varphi\varphi}_{reg}(s)=(2\pi)^{1-d}r^{-4}\int d^{d-2}p\,\int\limits_0^\infty
 dk\,k\left({\textbf{p}}{\,}^2+k^2+m^2\right)^{-s-1}\sum_{n\in\mathbb Z}
\left(n-\Phi\right)^2J^2_{|n-\Phi|}(kr)-\\
 -r^{-2}\left[\frac14\triangle_r-\xi\partial^2_r\right]
{\tilde t}{\,}^{00}_{reg}(s+1)\,,\,\,
\end{multline}
\begin{multline}\label{a28}
 t^{jj}_{reg}(s)=(2\pi)^{1-d}\int d^{d-2}p\,\,
 \left(p^j\right)^2\int\limits_0^\infty
 dk\,k\left({\textbf{p}}^2+k^2+m^2\right)^{-s-1}\sum_{n\in\mathbb Z}
J^2_{|n-\Phi|}(kr)+\\
-\left(\frac14-\xi\right)\triangle_r {\tilde
t}{\,}^{00}_{reg}(s+1)\,,\quad j=\overline{3,d}\,,\,
\end{multline}
where $\triangle_r=\partial^2_r+r^{-1}\partial_r\,$ is the
transverse radial part of the laplacian and
\begin{equation}\label{a29}
 {\tilde t}{\,}^{00}_{reg}(s)=
 (2\pi)^{1-d}\int d^{d-2}p\,\int\limits_0^\infty
  dk\,k\left({\textbf{p}}^2+k^2+m^2\right)^{-s}
  \sum_{n\in\mathbb Z}J^2_{|n-\Phi|}(kr)\,.
\end{equation}

Defining the fractional part of the flux,
\begin{equation}\label{a30}
  F=\Phi-[\![\Phi]\!]\,,\quad 0\leq F < 1\,,
\end{equation}
where $[\![u]\!]$ is the integer part of quantity $u$ (i.e. the
integer which is less than or equal to $u$), one can note that
tensor components (\ref{a25})-(\ref{a28}) are periodic functions
of flux $\Phi$, since they depend only on $F$ (being symmetric
under $F \rightarrow 1-F $)\,.


\section{Regularized vacuum expectation value}
Let us start by considering quantity
\begin{equation}\label{b1}
   {\tilde t}{\,}^{jj}_{reg}(s)=
   \left.t{\,}^{jj}_{reg}(s)\right|_{\xi=1/4}\,.
\end{equation}
Performing the integration over $\textbf{p}$, one gets
\begin{equation}\label{b2}
  {\tilde t}{\,}^{jj}_{reg}(s)=
   \frac1{(4\pi)^{\frac d2}}\frac{\Gamma(s+1-\frac d2)}{\Gamma(s+1)}
   \int\limits_0^\infty dk\,\,k\left(k^2+m^2\right)^{\frac d2-s-1}
   {\Sigma}_0(kr)\,,
\end{equation}
where $\Gamma(z)$ is the Euler gamma function and
\begin{equation}\label{b3}
   {\Sigma}_0(kr)=\sum_{n\in\mathbb Z}J^2_{|n-\Phi|}(kr)\,.
\end{equation}
Using relation (see, e.g., Ref.\cite{Prud})
\begin{equation}\label{b4}
   \sum_{\genfrac{}{}{0pt}{}{n\in\mathbb Z}{n\geq1}}J^2_{n+\mu}(z)=
   \mu\int\limits_0^z \frac{d\tau}\tau J^2_\mu(\tau)-\frac 12J^2_\mu(z)\,,\quad
   \mu>0\,,
\end{equation}
the summation over $n$ in Eq.(\ref{b3}) is performed in the case
of $\Phi\neq n$\,,
\begin{equation}\label{b5}
 {\Sigma}_0(kr)=\int\limits_0^{kr}
 d\tau\,\left[J_F(\tau)J_{-1+F}(\tau)+J_{-F}(\tau)J_{1-F}(\tau)\right]\,,\quad 0<F<1\,.
\end{equation}
Thus, Eq.(\ref{b2}) takes form (after integration by parts):
\begin{multline}\label{b6}
  {\tilde t}{\,}^{jj}_{reg}(s)=\frac r{2(4\pi)^{\frac d2}}
  \frac{\Gamma(s-\frac d2)}{\Gamma(s+1)}
  \int\limits_0^\infty dk\, \left(k^2+m^2\right)^{\frac d2-s}\times\\
  \times\left[J_F(kr)J_{-1+F}(kr)+J_{-F}(kr)J_{1-F}(kr)\right]\,.
\end{multline}
Using relations
\begin{equation}\label{b7}
  \left(k^2+m^2\right)^{-z}=\frac1{\Gamma(z)}
  \int\limits_0^\infty
  dy\,y^{z-1}\mathrm{exp}\left[-y\left(k^2+m^2\right)\right]\,,
  \qquad \mathrm{Re}\,z >0\,,
\end{equation}
and
\begin{eqnarray}\label{b8}
 \int\limits_0^\infty dk\,e^{-yk^2}J_\mu(kr)J_{\mu-1}(kr)=
 \frac1{2r}\int\limits_0^{r^2(2y)^{-1}}du\,e^{-u}
 [I_{\mu-1}(u)-I_\mu(u)]\,,\qquad \mu > 0\,,
\end{eqnarray}
 where $I_\mu(u)$ is the modified Bessel function of order
 $\mu$, we get
\begin{eqnarray}\label{b9}
  {\tilde t}{\,}^{jj}_{reg}(s)=
  \frac{2\sin(F\pi)}{(4\pi)^{\frac d2+1}}
  \frac{m^{d-2s}}{\Gamma(s+1)}
  \int\limits_0^\infty du\,e^{-u}
  [K_F(u)+K_{1-F}(u)]
  \gamma\left(s-\frac d2,\frac{m^2r^2}{2u}\right)\,,
\end{eqnarray}
where
$$
K_\mu(u)=\frac{\pi}{2\sin(\mu\pi)}[I_{-\mu}(u)-I_\mu(u)]
$$
is the Macdonald function of order $\mu$ and
$$
\gamma(z,w)=\int\limits_0^w d\tau\,\tau^{z-1}e^{-\tau}
$$
is the incomplete gamma function. Although relation (\ref{b9})
has been derived at \mbox{Re\,$s > \frac d2-1$}, it can be
continued analytically to the whole complex $s$-plane.
Introducing complementary function
$$
\Gamma(z,w) = \Gamma(z)-\gamma(z,w)= \int\limits_w^\infty
d\tau\,\tau^{z-1}e^{-\tau}\,,
$$
and decomposing Eq.(\ref{b9}) appropriately into a sum of two
terms, we get
\begin{multline}\label{b10}
 {\tilde t}{\,}^{jj}_{reg}(s)=\frac{m^{d-2s}}{2(4\pi)^{\frac d2}}
 \frac{\Gamma\left(s-\frac d2\right)}{\Gamma(s+1)}-\\
 -\frac{2\sin(F\pi)}{(4\pi)^{\frac d2+1}}
  \frac{m^{d-2s}}{\Gamma(s+1)}
  \int\limits_0^\infty du\,e^{-u}
  [K_F(u)+K_{1-F}(u)]
  \Gamma\left(s-\frac d2,\frac{m^2r^2}{2u}\right)=\\
=\frac{m^{d-2s}}{2(4\pi)^{\frac d2}}
 \frac{\Gamma\left(s-\frac d2\right)}{\Gamma(s+1)}-
 \frac{8\sin(F\pi)}{(4\pi)^{\frac d2+1}\Gamma(s+1)}
\left(\frac mr\right)^{\frac d2 -s}\times\\
\times \int\limits_1^\infty \frac{d\upsilon}{\sqrt{\upsilon^2-1}}
\cosh[(2F-1)\,\mathrm {arccosh}\, \upsilon] \upsilon^{s-\frac d2
-1}K_{s-\frac d2}(2mrv)\,,
\end{multline}
where in deriving the last line we have used relations (see, e.g.,
Ref.\cite{Prud})
\begin{equation}\label{b11}
  K_\mu(u)=\frac 12 \int\limits_0^\infty
  d\tau\,\tau^{\mu-1}\mathrm{exp}\left[-\frac u2(\tau+\tau^{-1})\right]
\end{equation}
and
\begin{equation}\label{b12}
\int\limits_0^\infty
\frac{d\tau}{\tau^2}\,\mathrm{exp}\left(-\frac{p}{2\tau}
\right)\Gamma\left(q,\frac{c\tau}2\right)= \left(\frac p4
\right)^ {\frac{q-2}2}c^{\frac q2}K_q(\sqrt{pc})\,.
\end{equation}

In a similar way we get\footnote{Note that this quantity was
first computed in Ref.\cite{Sit}, where it is called as the zeta
function density.}
\begin{multline}\label{b13}
{\tilde t}{\,}^{00}_{reg}(s)=\frac{m^{d-2s}}{(4\pi)^{\frac d2}}
 \frac{\Gamma\left(s-\frac d2\right)}{\Gamma(s)}-\\
-\frac{16\sin(F\pi)}{(4\pi)^{\frac d2+1}\Gamma(s)} \left(\frac mr\right)^{\frac
d2 -s}  \int\limits_1^\infty \frac{d\upsilon}{\sqrt{\upsilon^2-1}}
\cosh[(2F-1)\, \mathrm{arccosh}\, \upsilon] \upsilon^{s-\frac d2 -1}K_{\frac
d2-s}(2mrv)\,.
\end{multline}

Then we derive relations
\begin{multline}\label{b14}
 r^{-1}\partial_r{\tilde t}{\,}^{00}_{reg}(s+1)=
-\frac{32\sin(F\pi)}{(4\pi)^{\frac d2+1}\Gamma(s+1)} \left(\frac
mr\right)^{\frac d2 -s}\times \\
\times  \int\limits_1^\infty
\frac{d\upsilon}{\sqrt{\upsilon^2-1}} \cosh[(2F-1)\,
\mathrm{arccosh}\, \upsilon] \upsilon^{1+s-\frac d2}K_{\frac
d2-s}(2mrv)\,,
\end{multline}
\begin{multline}\label{b15}
\partial^2_r{\tilde t}{\,}^{00}_{reg}(s+1)=
-\frac{32\sin(F\pi)}{(4\pi)^{\frac d2+1}\Gamma(s+1)} \left(\frac
mr\right)^{\frac d2 -s} \int\limits_1^\infty
\frac{d\upsilon}{\sqrt{\upsilon^2-1}} \cosh[(2F-1)\,
\mathrm{arccosh}\, \upsilon]\times\\
\times \upsilon^{1+s-\frac d2}\left[K_{\frac d2-s}(2mrv)-2mr\upsilon K_{\frac
d2-s+1}(2mr\upsilon)\right]\,.
\end{multline}

Consequently, temporal and longitudinal components of the
regularized vacuum  energy-momentum tensor, Eqs.(\ref{a25}) and
(\ref{a28}), are given by expressions:
\begin{multline}\label{b16}
  t^{00}_{reg}(s)= \frac{m^{d-2s}}{(4\pi)^{\frac d2}}
 \frac{\Gamma\left(s-\frac d2\right)}{\Gamma(s)}+\\
 +\frac{16\sin(F\pi)}{(4\pi)^{\frac d2+1}\Gamma(s+1)} \left(\frac
mr\right)^{\frac d2 -s}\int\limits_1^\infty
\frac{d\upsilon}{\sqrt{\upsilon^2-1}} \cosh[(2F-1)\,
\mathrm{arccosh}\, \upsilon]\times\\
\times \upsilon^{s-\frac d2-1}\left\{
[-s+(1-4\xi)\upsilon^2]K_{\frac d2-s}(2mrv)-(1-4\xi)mr\upsilon^3
K_{\frac d2-s+1}(2mr\upsilon)\right\}\,,
\end{multline}
\begin{multline}\label{b17}
t^{jj}_{reg}(s)= \frac{m^{d-2s}}{(4\pi)^{\frac d2}}
 \frac{\Gamma\left(s-\frac d2\right)}{\Gamma(s+1)}-\\
 -\frac{16\sin(F\pi)}{(4\pi)^{\frac d2+1}\Gamma(s+1)} \left(\frac
mr\right)^{\frac d2 -s}\int\limits_1^\infty
\frac{d\upsilon}{\sqrt{\upsilon^2-1}} \cosh[(2F-1)\,\mathrm
{arccosh}\, \upsilon]\times\\
\times \upsilon^{s-\frac d2-1}\left\{
\left[\frac12+(1-4\xi)\upsilon^2\right]K_{\frac
d2-s}(2mrv)-(1-4\xi)mr\upsilon^3 K_{\frac
d2-s+1}(2mr\upsilon)\right\}\,.
\end{multline}

Let us turn now to transverse components of the regularized vacuum
energy-momentum tensor. Integrating over $\textbf{p} $ in the
first lines of Eq.(\ref{a26}) and (\ref{a27}), we get
\begin{multline}\label{b18}
 t^{rr}_{reg}(s)=
\frac{\Gamma(s+2-\frac
 d2)}{(4\pi)^{\frac d2}\,\Gamma(s+1)}\left[\Omega_1(s)-m^2\Omega_1(s+1)\right]-\\
 -\left[\frac 14 \partial^2_r-\left(\xi-\frac
 14\right)r^{-1}\partial_r\right]{\tilde t}{\,}^{00}_{reg}(s+1)\,,
\end{multline}
\begin{multline}\label{b19}
 t^{\varphi\varphi}_{reg}(s)=
\frac{r^{-2}\,\Gamma(s+2-\frac
 d2)}{(4\pi)^{\frac d2}\,\Gamma(s+1)}\left[\Omega_2(s)-m^2\Omega_2(s+1)\right]-\\
 -r^{-2}\left[\frac 14 r^{-1}\partial_r-\left(\xi-\frac
 14\right)\partial^2_r\right]{\tilde t}{\,}^{00}_{reg}(s+1)\,,
\end{multline}
where

\begin{equation}\label{b20}
  \Omega_a(s)=2\int\limits_0^\infty dk\,\,k\left(k^2+m^2\right)^
  {\frac d2-s-1}{\Sigma}_a(kr)\,,\quad a=1,2\,,
\end{equation}
\begin{multline}\label{b21}
  {\Sigma}_1(kr)=k^{-2}\sum_{n\in\mathbb
  Z}\left[\partial_rJ_{|n-\Phi|}(kr)\right]^2=\\
  =\frac 14 \sum_{n\in\mathbb Z}\left[J_{|n-\Phi|+1}(kr)-
  J_{|n-\Phi|-1}(kr)\right]^2\,,
\end{multline}
and
\begin{multline}\label{b22}
  {\Sigma}_2(kr)=(kr)^{-2}\sum_{n\in\mathbb
  Z}\left(n-\Phi\right)^2J^2_{|n-\Phi|}(kr)=\\
  =\frac 14 \sum_{n\in\mathbb Z}\left[J_{|n-\Phi|+1}(kr)+
  J_{|n-\Phi|-1}(kr)\right]^2\,.
\end{multline}
Using Eq.(\ref{b4}) and relation \cite{Prud}
\begin{multline}\label{b23}
 \sum_{\genfrac{}{}{0pt}{}{n\in\mathbb Z}{n\geq1}}
 J_{n+\mu+1}(z)J_{n+\mu-1}(z)=
  \frac 12 \mu J^2_\mu(z)+\frac 12 (1+\mu)J^2_{-1+\mu}(z)-\\
  -\frac{\mu(1+\mu)}z J_\mu(z)J_{-1+\mu}(z)\,,
\end{multline}
we get
\begin{equation}\label{b24}
  {\Sigma}_a(kr)=\frac 12 \left[{\Sigma}_0(kr)+
  {\tilde\Sigma}_a(kr)\right]\,,\quad a=1,2\,\,,
\end{equation}
where
\begin{multline}\label{b25}
  {\tilde\Sigma}_1(kr)=\frac 12 (1+F)
  \left[J^2_{-F}(kr)-J^2_F(kr)\right]
  +\frac 12 (2-F)
  \left[J^2_{-1+F}(kr)-J^2_{1-F}(kr)\right]-\\
  -\frac{F(1-F)}{kr}
  \left[J_F(kr)J_{-1+F}(kr)+J_{-F}(kr)J_{1-F}(kr)\right]\,,
\end{multline}
\begin{multline}\label{b26}
  {\tilde\Sigma}_2(kr)=\frac 12 (1-F)
  \left[J^2_{-F}(kr)-J^2_F(kr)\right]
  +\frac 12 F
  \left[J^2_{-1+F}(kr)-J^2_{1-F}(kr)\right]+\\
  +\frac{F(1-F)}{kr}
  \left[J_F(kr)J_{-1+F}(kr)+J_{-F}(kr)J_{1-F}(kr)\right]\,,
\end{multline}
and ${\Sigma}_0(kr)$ is given by Eq.(\ref{b5}). Thus,
Eq.(\ref{b20}) takes form
\begin{equation}\label{b27}
  \Omega_a(s)=\Omega_0(s)+{\tilde\Omega}_a(s)\,,\quad a=1,2\,\,,
\end{equation}
where
\begin{equation}\label{b28}
  \Omega_0(s)=\frac r{2s-d}\int\limits_0^\infty dk\,
  \left(k^2+m^2\right)^{\frac d2 -s}
   \left[J_F(kr)J_{-1+F}(kr)+J_{-F}(kr)J_{1-F}(kr)\right]\,,
\end{equation}
\begin{equation}\label{b29}
  {\tilde\Omega}_a(s)=\int\limits_0^\infty dk\,k
  \left(k^2+m^2\right)^{\frac d2 -s-1}
  {\tilde\Sigma}_a(kr)\,,\quad a=1,2\,\,.
\end{equation}

The integral in Eq.(\ref{b28}) has been already encountered during
the analysis of ${\tilde t}{\,}^{jj}_{reg}(s)\,$, see
Eq.(\ref{b6}). Similarly, using relations (\ref{b7}), (\ref{b8})
and (see Ref.\cite{Prud})
\begin{equation}\label{b30}
\int\limits_0^\infty dk\,k\,e^{-yk^2}J^2_\mu(kr)= \frac 1{2r^2}
\mathrm{exp}\left(-\frac{r^2}{2y}\right)I_\mu\left(\frac{r^2}{2y}\right)\,,\,
\mu>-1\,,
\end{equation}
we get
\begin{multline}\label{b31}
 {\tilde\Omega}_1(s)=\frac{\sin(F\pi)}{\pi\Gamma\left(s+1-\frac
 d2\right)}
 \left\{\frac 14\left(\frac{r^2}2\right)^{s-\frac d2}
 \int\limits_0^\infty d\tau\,\tau^{\frac d2 -s-2}
 \mathrm{exp}\left(-\tau-\frac{m^2r^2}{2\tau}\right)\right.\times\\
\times [(1+F)K_F(\tau)+(2-F)K_{1-F}(\tau)]-
F(1-F)r^{-2}m^{d-2s-2}\times \\
 \times \left.
 \int\limits_0^\infty du\,e^{-u}
 [K_F(u)+K_{1-F}(u)]
\gamma\left(1+s-\frac d2 , \frac{m^2r^2}{2u}\right)\right\}\,,
\end{multline}
\begin{multline}\label{b32}
{\tilde\Omega}_2(s)=\frac{\sin(F\pi)}{\pi\Gamma\left(s+1-\frac
 d2\right)}
 \left\{\frac 14\left(\frac{r^2}2\right)^{s-\frac d2}
 \int\limits_0^\infty d\tau\,\tau^{\frac d2 -s-2}
 \mathrm{exp}\left(-\tau-\frac{m^2r^2}{2\tau}\right)\right.\times\\
\times [(1-F)K_F(\tau)+FK_{1-F}(\tau)]+
F(1-F)r^{-2}m^{d-2s-2}\times \\
 \times \left.
 \int\limits_0^\infty du\,e^{-u}
 [K_F(u)+K_{1-F}(u)]
\gamma\left(1+s-\frac d2 , \frac{m^2r^2}{2u}\right)\right\}\,.
\end{multline}

Although relations (\ref{b31}) and (\ref{b32}) have been derived
at Re $s > \frac d2 -1$, they can be continued analytically to the
whole complex $s$-plane. Using integration by parts in
$\tau$-integrals and relations (\ref{b11}), (\ref{b12}) and
\begin{multline}\label{b33}
e^{-\tau}K_{\mu}(\tau)=2\int\limits_1^\infty d\upsilon
\left\{\frac{\upsilon}{\sqrt{\upsilon^2-1}}\cosh[(2\mu-1)\,\mathrm
{arccosh}\,\upsilon]+\right.\\
\left.\vphantom{\frac{\upsilon}{\sqrt{\upsilon^2-1}}}
 +\sinh[(2\mu-1)\,\mathrm
{arccosh}\,\upsilon]\right\}e^{-2\tau\upsilon^2}\,,
\end{multline}
 we get
\begin{multline}\label{b34}
  {\tilde\Omega}_1(s)-m^2 {\tilde\Omega}_1(s+1)=\\
=\frac{4\sin(F\pi)}{\pi\Gamma(2+s-\frac d2)} \left(\frac
mr\right)^{\frac d2 -s} \int\limits_1^\infty
\frac{d\upsilon}{\sqrt{\upsilon^2-1}} \cosh[(2F-1)\,\mathrm
{arccosh}\, \upsilon]\times\\
\times \upsilon^{s-\frac d2}\left[\upsilon K_{\frac
d2-s}(2mrv)-\frac12 mr(1+2\upsilon^2) K_{\frac
d2-s+1}(2mr\upsilon)\right]\,,
\end{multline}
\begin{multline}\label{b35}
  {\tilde\Omega}_2(s)-m^2 {\tilde\Omega}_2(s+1)=\\
=\frac{4\sin(F\pi)}{\pi\Gamma(2+s-\frac d2)} \left(\frac
mr\right)^{\frac d2 -s} \int\limits_1^\infty
\frac{d\upsilon}{\sqrt{\upsilon^2-1}} \cosh[(2F-1)\,\mathrm
{arccosh}\, \upsilon]\times\\
\times \upsilon^{s-\frac d2}\left[\upsilon K_{\frac
d2-s}(2mrv)+\frac12 mr(1-2\upsilon^2) K_{\frac
d2-s+1}(2mr\upsilon)\right]\,.
\end{multline}

Thus, we obtain following expressions for transverse components of the
regularized vacuum energy-momentum tensor, Eqs.(\ref{a26}) and (\ref{a27}):
\begin{multline}\label{b36}
t^{rr}_{reg}(s)= \frac{m^{d-2s}}{2(4\pi)^{\frac d2}}
 \frac{\Gamma\left(s-\frac d2\right)}{\Gamma(s+1)}-\\
-\frac{8\sin(F\pi)}{(4\pi)^{\frac d2+1}\Gamma(s+1)} \left(\frac
mr\right)^{\frac d2 -s}\int\limits_1^\infty
\frac{d\upsilon}{\sqrt{\upsilon^2-1}} \cosh[(2F-1)\,\mathrm
{arccosh}\, \upsilon]\times\\
\times \upsilon^{s-\frac d2-1} (1-4\xi\upsilon^2)K_{\frac
d2-s}(2mrv)\,,
\end{multline}
\begin{multline}\label{b37}
t^{\varphi\varphi}_{reg}(s)= \frac{r^{-2}m^{d-2s}}{2(4\pi)^{\frac
d2}}
 \frac{\Gamma\left(s-\frac d2\right)}{\Gamma(s+1)}-\\
-\frac{8\sin(F\pi)}{(4\pi)^{\frac d2+1}\Gamma(s+1)}r^{-2}
\left(\frac mr\right)^{\frac d2 -s}\int\limits_1^\infty
\frac{d\upsilon}{\sqrt{\upsilon^2-1}} \cosh[(2F-1)\,\mathrm
{arccosh}\, \upsilon]\times\\
\times \upsilon^{s-\frac d2-1} (1-4\xi\upsilon^2)[K_{\frac
d2-s}(2mrv)-2mr\upsilon K_{\frac d2-s+1}(2mrv)]\,.
\end{multline}

If, instead of Eq.(\ref{b7}), we use relation
\begin{equation}\label{b38}
  \left(k^2+m^2\right)^{-z}=\frac{2\sin(z\pi)}\pi
  \int\limits_0^\infty dy\,\frac{y^{1-2z}}{k^2+m^2+y^2}\,,\quad 0<\mathrm{Re}\,z<1\,,
\end{equation}
then we get components of the regularized vacuum energy-momentum
tensor in the following representation (see Appendix A):
\begin{multline}\label{b39}
 t^{00}_{reg}(s)=\frac{m^{d-2s}}{(4\pi)^{\frac d2}}
 \frac{\Gamma(s-\frac d2)}{\Gamma(s)}\,-\\
 -\frac{16\sin(F\pi)r^{2s-d}}{(4\pi)^{\frac d2+1}\Gamma(s+1)}
 \left\{\frac s{\Gamma(\frac d2-s+1)}
  \int\limits_{mr}^\infty dw\,\left(w^2-m^2r^2\right)^{\frac
 d2-s}K_F(w)K_{1-F}(w)+\right. \\
 \left.+\frac{1-4\xi}{2\,\Gamma(\frac d2-s-1)}
 \int\limits_{mr}^\infty dw\,\,w^3 \left(w^2-m^2r^2\right)^{\frac
 d2-s-2}\left[K^2_F(w)+K^2_{1-F}(w)\right]\right\}\,,
\end{multline}
\begin{multline}\label{b40}
 t^{jj}_{reg}(s)=\frac{m^{d-2s}}{2(4\pi)^{\frac d2}}
 \frac{\Gamma(s-\frac d2)}{\Gamma(s+1)}\,-\\
 -\frac{8\sin(F\pi)r^{2s-d}}{(4\pi)^{\frac d2+1}\Gamma(s+1)}
 \left\{\frac 1{\Gamma(\frac d2-s+1)}
  \int\limits_{mr}^\infty dw\,\left(w^2-m^2r^2\right)^{\frac
 d2-s}K_F(w)K_{1-F}(w)-\right. \\
 \left.-\frac{1-4\xi}{\,\Gamma(\frac d2-s-1)}
 \int\limits_{mr}^\infty dw\,\,w^3 \left(w^2-m^2r^2\right)^{\frac
 d2-s-2}\left[K^2_F(w)+K^2_{1-F}(w)\right]\right\}\,,
 \end{multline}
\begin{multline}\label{b41}
 t^{rr}_{reg}(s)=\frac{m^{d-2s}}{2(4\pi)^{\frac d2}}
 \frac{\Gamma \left(s-\frac d2\right)}{\Gamma(s+1)}-\\
 -\frac{8\sin (F\pi) r^{2s-d}}{(4\pi)^{\frac d2+1}\Gamma(s+1)}\left\{
 \frac{1}{\Gamma\left(\frac d2 -s+1\right)}
 \int\limits^\infty_{mr} dw
\left(w^2-m^2r^2\right)^{\frac d2-s}\right.\times\\
\times\left\{
K_F(w)K_{1-F}(w)-\frac w2 \left[K_F^2(w)+K_{1-F}^2(w)\right]\right\}+\\
 +\frac{1}{\Gamma\left(\frac d2 -s-1\right)}
 \int\limits^\infty_{mr} dw\,w^2
  \left(w^2-m^2r^2\right)^{\frac d2-s-2}
\left\{2[F(1-F)-2\xi] K_F(w)K_{1-F}(w)+\right.\\
\left.\left.+w
  \left[F K^2_F(w)+(1-F)
  K^2_{1-F}(w)\right]\,\right\}
  \vphantom{\int\limits^\infty_{mr}}\right\}\,,
\end{multline}
\begin{multline}\label{b42}
 t^{\varphi\varphi}_{reg}(s)=\frac{r^{-2}m^{d-2s}}{2(4\pi)^{\frac d2}}
 \frac{\Gamma \left(s-\frac d2\right)}{\Gamma(s+1)}-\\
 -\frac{8\sin (F\pi) r^{2s-d-2}}{(4\pi)^{\frac d2+1}\Gamma(s+1)}
\left\{ \frac{1}{\Gamma\left(\frac d2 -s+1\right)}
 \int\limits^\infty_{mr} dw
\left(w^2-m^2r^2\right)^{\frac d2-s}\right.\times\\
\times\left\{
K_F(w)K_{1-F}(w)-\frac w2 \left[K_F^2(w)+K_{1-F}^2(w)\right]\right\}-\\
-\frac{1}{\Gamma\left(\frac d2 -s-1\right)}
 \int\limits^\infty_{mr} dw\,w^2
  \left(w^2-m^2r^2\right)^{\frac d2-s-2}
\left\{2[F(1-F)-2\xi] K_F(w)K_{1-F}(w)+\right.\\
 \left.\left.+w
  \left[(F-4\xi) K^2_F(w)+(1-F-4\xi)
  K^2_{1-F}(w)\right]\,\right\}
  \vphantom{\int\limits^\infty_{mr}}\right\}\,.
\end{multline}

Whereas expressions (\ref{b16}), (\ref{b17}), (\ref{b36}) and (\ref{b37})are
defined on the whole complex s-plane, expressions (\ref{b39}) - (\ref{b42}) are
defined on a half-plane Re\,$s < d/2 - 1$.


\section{Renormalized vacuum expectation value}
In the absence of the $d-2$ - brane (i.e. at $\Phi=0$) expressions
(\ref{b16}), (\ref{b17}), (\ref{b36}) and (\ref{b37}) (or
(\ref{b39})-(\ref{b42}) take form
$$
t^{00}_{reg}(s)|_{F=0}=\frac{m^{d-2s}}{(4\pi)^{\frac d2}}
 \frac{\Gamma(s-\frac d2)}{\Gamma(s)}\,,
$$
\begin{equation}\label{c1}
t^{jj}_{reg}(s)|_{F=0}=t^{rr}_{reg}(s)|_{F=0}=
 r^2t^{\varphi\varphi}_{reg}(s)|_{F=0}=
\frac{m^{d-2s}}{2(4\pi)^{\frac d2}}
 \frac{\Gamma(s-\frac d2)}{\Gamma(s+1)}\,.
\end{equation}
These quantities are eliminated by the requirement of the normal ordering of
the operator product in the case of noninteracting quantized field. For
consistency with the noninteracting case, one has to subtract quantities
(\ref{c1}) from the regularized expressions corresponding to the interaction
with the background. Then, taking limit $s\rightarrow -\frac 12$, one defines
renormalized vacuum energy-momentum tensor:
\begin{equation}\label{c2}
t^{\mu\nu}_{ren}= \lim_{s\rightarrow -\frac 12}
[t^{\mu\nu}_{reg}(s)-t^{\mu\nu}_{reg}(s)|_{F=0}],
\end{equation}
which is of physical interest. Its components are given by
expressions:
\begin{multline}\label{c3}
t^{00}_{ren}= -t^{jj}_{ren}=
\frac{16\sin(F\pi)}{(4\pi)^{\frac{d+3}2}} \left(\frac
mr\right)^{\frac{d+1}2}\int\limits_1^\infty
\frac{d\upsilon}{\sqrt{\upsilon^2-1}} \cosh[(2F-1)\,\mathrm
{arccosh}\, \upsilon]\times\\
\times \upsilon^{-\frac{d+3}2}\left\{
[1+2(1-4\xi)\upsilon^2]K_{\frac{d+1}2}(2mrv)-2(1-4\xi)mr\upsilon^3
K_{\frac{d+3}2}(2mr\upsilon)\right\}\,,
\end{multline}
\begin{multline}\label{c4}
t^{rr}_{ren}= - \frac{16\sin(F\pi)}{(4\pi)^{\frac{d+3}2}}
\left(\frac mr\right)^{\frac{d+1}2}\int\limits_1^\infty
\frac{d\upsilon}{\sqrt{\upsilon^2-1}} \cosh[(2F-1)\,\mathrm
{arccosh}\, \upsilon]\times\\
\times \upsilon^{-\frac{d+3}2}(1-4\xi\upsilon^2)
K_{\frac{d+1}2}(2mr\upsilon)\,,
\end{multline}
\begin{multline}\label{c5}
t^{\varphi\varphi}_{ren}= -
\frac{16\sin(F\pi)}{(4\pi)^{\frac{d+3}2}}\frac1{r^2}\left(\frac
mr\right)^{\frac{d+1}2}\int\limits_1^\infty
\frac{d\upsilon}{\sqrt{\upsilon^2-1}} \cosh[(2F-1)\,\mathrm
{arccosh}\, \upsilon]\times\\
\times \upsilon^{-\frac{d+3}2}(1-4\xi\upsilon^2)
\left\{K_{\frac{d+1}2}(2mr\upsilon)-2mr\upsilon
K_{\frac{d+3}2}(2mr\upsilon)\right\}\,,
\end{multline}
or, in the alternative representation,
\begin{multline}\label{c6}
 t^{00}_{ren}=-t^{jj}_{ren}=
 \frac{16\sin(F\pi)}{(4\pi)^{\frac{d+3}2}}r^{-d-1}
 \left\{\frac 1{\Gamma(\frac{d+3}2)}
  \int\limits_{mr}^\infty dw\,\left(w^2-m^2r^2\right)^{\frac{d+1}2}
  K_F(w)K_{1-F}(w)-\right. \\
 \left.-\frac{1-4\xi}{\Gamma(\frac{d-1}2)}
 \int\limits_{mr}^\infty dw\,\,w^3
 \left(w^2-m^2r^2\right)^{\frac{d-3}2}
 \left[K^2_F(w)+K^2_{1-F}(w)\right]\right\}\,,
\end{multline}
\begin{multline}\label{c7}
 t^{rr}_{ren}=
 -\frac{16\sin(F\pi)}{(4\pi)^{\frac{d+3}2}}r^{-d-1}
 \left\{\frac 1{\Gamma(\frac{d+3}2)}
  \int\limits_{mr}^\infty dw\,\left(w^2-m^2r^2\right)^{\frac{d+1}2}\times\right. \\
  \times\left\{K_F(w)K_{1-F}(w)-\frac
  w2\left[K^2_F(w)+K^2_{1-F}\right]\right\}
+\frac1{\Gamma(\frac{d-1}2)}
 \int\limits_{mr}^\infty dw\,\,w^2
 \left(w^2-m^2r^2\right)^{\frac{d-3}2}\times\\
\left.\times \left\{2\,[F(1-F)-2\xi]\,K_F(w)K_{1-F}(w)
+w\left[F\,K^2_F(w)+(1-F)\,K^2_{1-F}(w)\right]\right\}
\vphantom{\int\limits_{mr}^\infty}\right\}\,,
\end{multline}
\begin{multline}\label{c8}
 t^{\varphi\varphi}_{ren}=
 -\frac{16\sin(F\pi)}{(4\pi)^{\frac{d+3}2}}r^{-d-3}\times\\ \times
 \left\{\frac 1{\Gamma(\frac{d+3}2)}
  \int\limits_{mr}^\infty dw\,\left(w^2-m^2r^2\right)^{\frac{d+1}2}\left\{K_F(w)K_{1-F}(w)-\frac
  w2\left[K^2_F(w)+K^2_{1-F}\right]\right\}-\right.\\
-\frac1{\Gamma(\frac{d-1}2)}
 \int\limits_{mr}^\infty dw\,\,w^2
 \left(w^2-m^2r^2\right)^{\frac{d-3}2}
\left\{2\,[F(1-F)-2\xi]\,K_F(w)K_{1-F}(w)+\right.\\
\left.\left.+w\left[(F-4\xi)\,K^2_F(w)+(1-F-4\xi)\,K^2_{1-F}(w)\right]\right\}
\vphantom{\int\limits_{mr}^\infty}\right\}\,.
\end{multline}

One can verify that relation
\begin{equation}\label{c9}
(\partial_r+r^{-1})t^{rr}_{ren}-r\,t^{\varphi\varphi}_{ren}=0\,,
\end{equation}
is valid; consequently, the vacuum energy-momentum tensor is
conserved:
\begin{equation}\label{c10}
\nabla_\mu t^{\mu\nu}_{ren}=0\,.
\end{equation}

Taking trace of the tensor, we get
\begin{multline}\label{c11}
g_{\mu\nu}t^{\mu\nu}_{ren}=
\frac{32\sin(F\pi)}{(4\pi)^{\frac{d+3}2}} \int\limits_1^\infty
\frac{d\upsilon}{\sqrt{\upsilon^2-1}} \cosh[(2F-1)\,\mathrm
{arccosh}\, \upsilon]\,\upsilon^{-\frac{d+1}2} \times\\
\times  \left\{(d-1-4\xi d)\upsilon\left(\frac
mr\right)^{\frac{d+1}2}
\left[K_{\frac{d+1}2}(2mr\upsilon)-mr\upsilon
K_{\frac{d+3}2}(2mr\upsilon)\right]-\right.\\
\left.-m^2 \left(\frac
mr\right)^{\frac{d-1}2}K_{\frac{d-1}2}(2mr\upsilon)\right\}\,.
\end{multline}
As  follows from the last relation, the trace becomes
proportional to the mass squared in the case of $\xi=\xi_c$
(\ref{intr1}), which is in accordance with general relation
(\ref{a10}).

To conclude this Section, we present the vacuum energy-momentum tensor in the
Cartesian coordinate frame:

\begin{equation}\label{c12}
\left(
\begin{array}{ccc}

\begin{array}{ccc}
\varepsilon & 0 & 0\\
0 & P^1 &
\frac 12 \left(P^1-P^2\right)
\frac{x^1\,x^2}{\left(x^1\right)^2-\left(x^2\right)^2}\\
0 & \frac 12
\left(P^1-P^2\right)\frac{x^1\,x^2}{\left(x^1\right)^2-\left(x^2\right)^2}
& P^2
\end{array} &
\begin{array}{c}
\mid \\ \mid \\ \mid \end{array} & 0\\

--------------------- \!\!\!\!\!\!\! & -\mid- & \!\!\!\!\!\!\!------ \\

0 &
\begin{array}{c}
\mid \\ \mid \\ \mid \end{array} &
\begin{array}{ccc}
P^3 &  & 0\\
 & \ddots & \\
0 &  & P^d
\end{array}

\end{array}
\right)\,,
\end{equation}
where
\begin{equation}\label{c13}
\begin{array}{l}
\varepsilon=-P^j=t^{00}_{ren}\,,\quad j=\overline{3,d}\,,\\
P^1=\frac{\left(x^1\right)^2}{\left(x^1\right)^2+\left(x^2\right)^2}t^{rr}_{ren}+
\left(x^2\right)^2t^{\varphi\varphi}_{ren}\,,\\
P^2=\frac{\left(x^2\right)^2}{\left(x^1\right)^2+\left(x^2\right)^2}t^{rr}_{ren}+
\left(x^1\right)^2t^{\varphi\varphi}_{ren}\,.
\end{array}
\end{equation}


\section{Asymptotics at large and small distances from the brane}
Components of the vacuum energy-momentum tensor depend on the
distance from the brane in the transverse direction. Using
representation (\ref{c6})-(\ref{c8}), it is straightforward to
determine the large distance behaviour of the tensor components:
\begin{multline}\label{f1}
t^{00}_{ren}=-\frac{2\sin(F\pi)}{(4\pi)^{\frac{d+1}2}}\,e^{-2mr}
\left(\frac{m}{r}\right)^{\frac{d+1}2} \left(\vphantom{\left(\frac
12\right)^34}1-4\xi-\right.\\
\left.-\left\{\frac12-(1-4\xi)
\left[\left(\frac{d+2}4\right)^2-\frac5{16}-F(1-F)\right]\right\}
(mr)^{-1}+\mathrm{O}\left[(mr)^{-2}\right]\right)\,,\,\,mr\gg1\,,
\end{multline}
\begin{multline}\label{f2}
t^{rr}_{ren}=-\frac{\sin(F\pi)}{(4\pi)^{\frac{d+1}2}}\,e^{-2mr}
\left(\frac{m}{r}\right)^{\frac{d-1}2}r^{-2}
\left(\vphantom{\left(\frac
12\right)^34}1-4\xi-\right.\\
\left.-\left\{\frac12-(1-4\xi)
\left[\left(\frac{d}4\right)^2+\frac3{16}-F(1-F)\right]\right\}
(mr)^{-1}+\mathrm{O}\left[(mr)^{-2}\right]\right)\,,\,\,mr\gg1\,,
\end{multline}
\begin{multline}\label{f3}
t^{\varphi\varphi}_{ren}=\frac{2\sin(F\pi)}{(4\pi)^{\frac{d+1}2}}\,e^{-2mr}
\left(\frac{m}{r}\right)^{\frac{d+1}2}r^{-2}
\left(\vphantom{\left(\frac
12\right)^34}1-4\xi-\right.\\
\left.-\left\{\frac12-(1-4\xi)
\left[\left(\frac{d+2}4\right)^2+\frac3{16}-F(1-F)\right]\right\}
(mr)^{-1}+\mathrm{O}\left[(mr)^{-2}\right]\right)\,,\,\,mr\gg1\,.
\end{multline}
The small-distance behaviour is given by expressions:
\begin{multline}\label{f4}
t^{00}_{ren}=-\frac{\sin(F\pi)}{(4\pi)^{\frac d2 +1}}
\frac{\Gamma\left(\frac{d+1}2-F\right)\Gamma\left(\frac{d-1}2+F\right)}
{\Gamma\left(\frac d2+1\right)}\times\\
\times\left[(d-1)^2-4\frac{F(1-F)}{d+1}-4\xi d(d-1)\right]r^{-d-1}
\left\{1+\mathrm{O}\left[(mr)^2\right]\right\},\,\,mr\ll 1\,,
\end{multline}
\begin{multline}\label{f5}
t^{rr}_{ren}=-\frac{\sin(F\pi)}{(4\pi)^{\frac d2 +1}}
\frac{\Gamma\left(\frac{d+1}2-F\right)\Gamma\left(\frac{d-1}2+F\right)}
{\Gamma\left(\frac d2+1\right)}\times\\
\times\left[d-1+4\frac{F(1-F)}{d+1}-4\xi d\right]r^{-d-1}
\left\{1+\mathrm{O}\left[(mr)^2\right]\right\},\,\,mr\ll 1\,,
\end{multline}
\begin{multline}\label{f6}
t^{\varphi\varphi}_{ren}=\frac{\sin(F\pi)}{(4\pi)^{\frac d2 +1}}
\frac{\Gamma\left(\frac{d+1}2-F\right)\Gamma\left(\frac{d-1}2+F\right)}
{\Gamma\left(\frac d2+1\right)}\times\\
\times d\left[d-1+4\frac{F(1-F)}{d+1}-4\xi d\right]r^{-d-3}
\left\{1+\mathrm{O}\left[(mr)^2\right]\right\},\,\,mr\ll 1\,.
\end{multline}
 Evidently, leading terms in the expressions in Eqs.(\ref{f4})-(\ref{f6})
 yield us the vacuum energy-momentum tensor in
the strictly massless case ($m=0$). We see that in this case the
tensor components are characterized by a simple power dependence
on the distance from the brane.


\section{Half-integer values of the brane flux}
As has been already noted, the vacuum energy-momentum tensor is periodic in the
value of the brane flux (i.e. depends on its fractional value only), vanishing
at its integer values ($F=0$) and being symmetric under change $F\rightarrow
1-F$. Moreover, as follows from Eqs.(\ref{c3})-(\ref{c5}) (or
(\ref{c6})-(\ref{c8})), maximal absolute values of the tensor components are
achieved at half-integer values of the brane flux ($F=\frac12$).

 Using representation (\ref{c3}) - (\ref{c5}) (see, e.g., first line in Eq.(\ref{b10})), we get
\begin{multline}\label{d1}
\left.t^{00}_{ren}\right|_{F=\frac12}=
\frac{2m^{d+1}}{(4\pi)^{\frac d2 +1}}\left\{\int\limits_0^\infty
d\tau\,\tau^{-\frac12}e^{-\tau}\Gamma\left(-\frac{d+1}2
,\frac{m^2r^2}\tau \right)+\right.\\
\left.+2(1-4\xi)(mr)^{-\frac d2 -1} \left[K_{\frac
d2}(2mr)-2mr\,K_{\frac
d2+1}(2mr)\right]\vphantom{\int\limits_0^\infty}\right\}\,,
\end{multline}
\begin{multline}\label{d2}
\left.t^{rr}_{ren}\right|_{F=\frac12}=
-\frac{2m^{d+1}}{(4\pi)^{\frac d2 +1}}\left\{\int\limits_0^\infty
d\tau\,\tau^{-\frac12}e^{-\tau}\Gamma\left(-\frac{d+1}2
,\frac{m^2r^2}\tau \right)-\right.\\
\left.-8\xi(mr)^{-\frac d2 -1} K_{\frac d2}(2mr)
\vphantom{\int\limits_0^\infty}\right\}\,,
\end{multline}
\begin{multline}\label{d3}
\left.t^{\varphi\varphi}_{ren}\right|_{F=\frac12}=
-\frac{2r^{-2}m^{d+1}}{(4\pi)^{\frac d2
+1}}\left\{\int\limits_0^\infty
d\tau\,\tau^{-\frac12}e^{-\tau}\Gamma\left(-\frac{d+1}2
,\frac{m^2r^2}\tau \right)-\right.\\
\left.-4(1-4\xi)(mr)^{-\frac d2} K_{\frac d2+1}(2mr)
\vphantom{\int\limits_0^\infty}\right\}\,.
\end{multline}

In Appendix B we show that integration in Eqs.(\ref{d1}) - (\ref{d3}) can be
performed in the case of physical values of space dimension $(d\in\mathbb Z,
d\geq2)$. As a result, quantities (\ref{d1}) - (\ref{d3}) are expressed in
terms of  Macdonald function $K_{\mu}(u)$ and modified Struve function
$L_{\mu}(u)$ of integer order in the case of even $d$, and in terms of the
integral exponential function $E_1(u)$ (see, e.g., Ref.\cite{Abra}) and
elementary functions in the case of odd $d$. In particular, we get
\begin{multline}\label{d4}
\left.t^{00}_{ren}\right|_{F=\frac12}=
\frac{m^3}{3\pi^2}\left\{\frac \pi2 - \pi mr
\left[K_0(2mr)L_{-1}(2mr)+K_1(2mr)L_0(2mr)\right]- \right.\\
\left.-(1-6\xi)(mr)^{-1}K_0(2mr)-\left[1+\frac12\left(\frac12-6\xi\right)
(mr)^{-2}\right]K_1(2mr)\right\}\,,\quad d=2\,,
\end{multline}
\begin{multline}\label{d5}
\left.t^{rr}_{ren}\right|_{F=\frac12}=
-\frac{m^3}{3\pi^2}\left\{\frac \pi2 - \pi mr
\left[K_0(2mr)L_{-1}(2mr)+K_1(2mr)L_0(2mr)\right]+ \right.\\
\left.+\frac12(mr)^{-1}K_0(2mr)-\left[1-\frac12(1-6\xi)(mr)^{-2}\right]
K_1(2mr)\right\}\,,\quad d=2\,,
\end{multline}
\begin{multline}\label{d6}
\left.t^{\varphi\varphi}_{ren}\right|_{F=\frac12}=
-\frac{r^{-2}m^3}{3\pi^2}\left\{\frac \pi2 - \pi mr
\left[K_0(2mr)L_{-1}(2mr)+K_1(2mr)L_0(2mr)\right]- \right.\\
\left.-(1-6\xi)(mr)^{-1}K_0(2mr)-[1+(1-6\xi)(mr)^{-2}]
K_1(2mr)\right\}\,,\quad d=2\,,
\end{multline}
\begin{multline}\label{d7}
\left.t^{00}_{ren}\right|_{F=\frac12}=
\frac{m^4}{(4\pi)^2}\left\{E_1(2mr)-\frac12
e^{-2mr}\left[(mr)^{-1}+\left(\frac72
-16\xi\right)(mr)^{-2}+\right.\right.\\
\left.\left.+\left(\frac52-16\xi\right)(mr)^{-3}+\frac12\left(\frac52-
16\xi\right)(mr)^{-4}\right]\right\}\,, \quad d=3\,,
\end{multline}
\begin{multline}\label{d8}
\left.t^{rr}_{ren}\right|_{F=\frac12}=
-\frac{m^4}{(4\pi)^2}\left\{E_1(2mr)-\frac12
e^{-2mr}\left[(mr)^{-1}-\frac12
(mr)^{-2}-\right.\right.\\
\left.\left.-\frac12(3-16\xi)(mr)^{-3}-\frac14(3-
16\xi)(mr)^{-4}\right]\right\}\,, \quad d=3\,,
\end{multline}
\begin{multline}\label{d9}
\left.t^{\varphi\varphi}_{ren}\right|_{F=\frac12}=
-\frac{r^{-2}m^4}{(4\pi)^2}\left\{E_1(2mr)-\frac12
e^{-2mr}\left[(mr)^{-1}+\left(\frac72
-16\xi\right)(mr)^{-2}+\right.\right.\\
\left.\left.+\frac32(3-16\xi)(mr)^{-3}+\frac34(3-
16\xi)(mr)^{-4}\right]\right\}\,, \quad d=3\,.
\end{multline}
Expressions in cases of arbitrary even and odd values of space dimension are
given in Appendix B (see Eqs.(\ref{ab6}) - (\ref{ab11})).


\section{The strong energy condition and its violation}
Energy-momentum tensor of the physically reasonable classical
matter satisfies the strong energy condition \cite{HaE}
\begin{equation}\label{e1}
T^{\mu\nu}u_{\mu}u_{\nu}-\frac 12 g_{\mu\nu}T^{\mu\nu}\geq0\,,
\end{equation}
where $u_{\mu}$ is a time-like vector $\left(u^{\mu}u_{\mu}=1\right)$. To
check, whether the vacuum energy-momentum tensor in the background of the
$d-2$-brane \footnote{In this Section we consider physical values of space
dimension: $d\geq2$\,.} satisfies condition (\ref{e1}), it is sufficient to
analyse three quantities: $t^{00}_{ren}-\frac 12 g_{\mu\nu}t^{\mu\nu}_{ren}$,
 $t^{00}_{ren}+t^{rr}_{ren}$,
$t^{00}_{ren}+r^2 t^{\varphi\varphi}_{ren}$.

In the massless case, using Eqs.(\ref{f4})-(\ref{f6}), we get
\begin{multline}\label{e2}
\left.\left(t^{00}_{ren}-\frac 12
g_{\mu\nu}t^{\mu\nu}_{ren}\right)\right|_{m=0}=
\frac{4\sin(F\pi)}{(4\pi)^{\frac d2+1}} \frac{\Gamma\left(\frac{d+1}2
-F\right)\Gamma\left(\frac{d-1}2
+F\right)}{\Gamma\left(\frac d2+1\right)}\times\\
\times\left[\frac12(d-2)(d-1)d(\xi_c-\xi)+\frac{F(1-F)}{d+1}\right]r^{-d-1}\,,
\end{multline}
\begin{multline}\label{e3}
\left.\left(t^{00}_{ren}+t^{rr}_{ren}\right)\right|_{m=0}=
-\frac{4\sin(F\pi)}{(4\pi)^{\frac d2+1}}
\frac{\Gamma\left(\frac{d+1}2 -F\right)\Gamma\left(\frac{d-1}2
+F\right)}{\Gamma\left(\frac d2+1\right)}\times\\
\times d^2(\xi_c-\xi)r^{-d-1}\,,
\end{multline}
\begin{multline}\label{e4}
\left.\left(t^{00}_{ren}+r^2
t^{\varphi\varphi}_{ren}\right)\right|_{m=0}=
\frac{4\sin(F\pi)}{(4\pi)^{\frac d2+1}}
\frac{\Gamma\left(\frac{d+1}2 -F\right)\Gamma\left(\frac{d-1}2
+F\right)}{\Gamma\left(\frac d2+1\right)}\times\\
\times [d(\xi_c-\xi)+F(1-F)]r^{-d-1}\,,
\end{multline}
where $\xi_c$ is given by Eq.(\ref{intr1}). All quantities
(\ref{e2})-(\ref{e4}) are simultaneously nonnegative at
\begin{equation}\label{e5}
\begin{array}{l}
\xi_c\leq\xi\leq\xi_c+\frac12 F(1-F)\,,\quad d=2\,,\\
\xi_c\leq\xi\leq\xi_c+\dfrac{2F(1-F)}{(d-2)(d-1)d(d+1)}\,,\quad d\geq3\,,
\end{array}
\end{equation}
then the strong energy condition is satisfied by the vacuum energy-momentum
tensor of the quantized massless scalar matter in the background of the brane.
If $F$ is infinitesimally close to 0 or to 1, then the condition is satisfied
only at $\xi=\xi_c$, i.e. when conformal invariance is maintained. Note, that
in the conformally invariant case the strong energy condition coincides, as a
consequence of the vanishing trace, with the weak energy one \cite{HaE}.

A similar analysis is carried out for the quantized massive scalar matter, and
we find that both the strong and weak energy conditions are violated  for all
values of $\xi$.

The temporal component of the vacuum tensor (i.e. energy density) is positive
at $\xi\geq\frac14$ and negative at $\xi\leq0$.\footnote{The vacuum energy
density at $\xi=\frac14$ was considered in Ref.\cite{Sit}\,.} Transverse
components of the vacuum tensor are also of opposite signs at $\xi\geq\frac14$
and at $\xi\leq0$: the radial one is of the same and the angular one is of the
opposite to the sign of the temporal component. The region $0<\xi<\frac14$ or,
more precisely, the vicinity of $\xi=\xi_c$ is distinguished as the region
where all components change their signs. Transverse components change their
signs at a certain value of $\xi$ simultaneously for all distances from the
brane. In contrast to this, the temporal component is positive at small
distances and negative at large distances for a certain, dependent on the value
of the brane flux, vicinity of the point $\xi=\xi_c$.

This situation is illustrated by Figures 1-3. Here variable $mr$ is along
$x$-axis, and dimensionless products of tensor components at half-integer
values of the brane flux, Eqs.(\ref{d4})-(\ref{d9}), and appropriate powers of
$r$ are along $y$-axis: $\left.r^{d+1}t^{00}_{ren}\right|_{F=\frac12}$ is
presented by a solid line, $\left.r^{d+1}t^{rr}_{ren}\right|_{F=\frac12}$ - by
a dotted line, and $\left.r^{d+3}t^{\varphi\varphi}_{ren}\right|_{F=\frac12}$ -
by a dashed line. We consider cases of $\xi=0$, $\xi=\xi_c$, $\xi=\frac14$,
each one at $d=2$ and $d=3$. We see from Fig.2 that the vacuum energy density
at $\xi=\xi_c$ has minimum at $mr\approx1.0\,\,(d=2)$ or
$mr\approx1.2\,\,(d=3)$; the minimal value is
$\left.r^{d+1}t^{00}_{ren}\right|_{F=\frac12}\approx-0.0018\,\,(d=2)$ or
$\left.r^{d+1}t^{00}_{ren}\right|_{F=\frac12}\approx-0.0039\,\,(d=3)$.

To conclude this Section, we present general expressions for the vacuum tensor
components in the case of conformal coupling:
\begin{multline}\label{e6}
\left.t^{00}_{ren}\right|_{\xi=\xi_c}= -
\frac{16\sin(F\pi)}{(4\pi)^{\frac{d+3}2}}\left(\frac
mr\right)^{\frac{d+1}2}\int\limits_1^\infty d\upsilon\,
\cosh[(2F-1)\,\mathrm
{arccosh}\, \upsilon]\upsilon^{-\frac{d+3}2}\times\\
\times
\left\{\sqrt{\upsilon^2-1}K_{\frac{d+1}2}(2mr\upsilon)+\frac{\upsilon^2}{d\sqrt{\upsilon^2-1}}
\left[2mr\upsilon\,
K_{\frac{d-1}2}(2mr\upsilon)-K_{\frac{d+1}2}(2mr\upsilon)\right]\right\}\,,
\end{multline}
\begin{multline}\label{e7}
\left.t^{rr}_{ren}\right|_{\xi=\xi_c}=
\frac{16\sin(F\pi)}{(4\pi)^{\frac{d+3}2}}\left(\frac
mr\right)^{\frac{d+1}2}\int\limits_1^\infty d\upsilon\,
\cosh[(2F-1)\,\mathrm
{arccosh}\, \upsilon]\upsilon^{-\frac{d+3}2}\times\\
\times
\left(\sqrt{\upsilon^2-1}-\frac{\upsilon^2}{d\sqrt{\upsilon^2-1}}\right)
K_{\frac{d+1}2}(2mr\upsilon)\,,
\end{multline}
\begin{multline}\label{e8}
\left.t^{\varphi\varphi}_{ren}\right|_{\xi=\xi_c}=-
\frac{16\sin(F\pi)}{(4\pi)^{\frac{d+3}2}}\frac1{r^2}\left(\frac
mr\right)^{\frac{d+1}2}\int\limits_1^\infty d\upsilon\,
\cosh[(2F-1)\,\mathrm
{arccosh}\, \upsilon]\upsilon^{-\frac{d+3}2}\times\\
\times
\left(\sqrt{\upsilon^2-1}-\frac{\upsilon^2}{d\sqrt{\upsilon^2-1}}\right)
\left[2mr\upsilon\,
K_{\frac{d-1}2}(2mr\upsilon)+d\,K_{\frac{d+1}2}(2mr\upsilon)\right]\,.
\end{multline}
Vacuum energy density (\ref{e6}) has minimum at $r=r_{min}$ where $r_{min}$ is
determined by equation
$$
\left.\left(\left.\frac d{d\,r}
t^{00}_{ren}\right|_{\xi=\xi_c}\right)\right|_{r=r_{min}}=0\,,
$$
or
\begin{multline}\label{e9}
\int\limits_1^\infty d\upsilon\, \cosh[(2F-1)\,\mathrm {arccosh}\,
\upsilon]\upsilon^{-\frac{d+1}2}\left\{\sqrt{\upsilon^2-1}\,
K_{\frac{d+3}2}(2mr_{min}\upsilon)+\right.\\
\left.+\frac{\upsilon^2}{d\sqrt{\upsilon^2-1}}\left[2mr_{min}\upsilon\,
K_{\frac{d+1}2}(2mr_{min}\upsilon)-K_{\frac{d+3}2}(2mr_{min}\upsilon)
\right]\right\}=0\,.
\end{multline}

\section{Summary}
We have shown that the vacuum of the quantized charged scalar
field is polarized in the background of a static magnetic $d-2$ -
brane in flat $d+1$ - dimensional space-time. Vector potential of
the brane induces a finite energy-momentum tensor in the vacuum;
therefore, this effect may be denoted as the
Casimir-Bohm-Aharonov effect. Tensor components depend
periodically on the brane flux $(\Phi)$, vanishing at its integer
values  $(\Phi=n)$, and attaining maximal absolute values at its
half-integer values $(\Phi=n+\frac12)$. A remarkable feature is a
possibility of analytic continuation in space dimension:
representation (\ref{c3})-(\ref{c5}) yields holomorphic functions
of $d$ on the whole complex $d$-plane, and representation
(\ref{c6})-(\ref{c8}) yields holomorphic functions of $d$ on a
half-plane Re $d>1$. The tensor components decrease exponentially
at large distances from the brane, see Eqs.(\ref{f1})-(\ref{f3}).
If the mass of scalar field is zero, then expressions for tensor
components are simplified considerably, see
Eqs.(\ref{f4})-(\ref{f6}).

Strong and weak energy conditions \cite{HaE} are usually violated
by vacuum polarization effects of quantized fields: in
particular, this happens in most cases of the conventional
Casimir effect \cite{Most}. However, we have found that these
conditions are satisfied by the vacuum energy-momentum tensor of
conformally invariant $(\xi=\xi_c)$ massless scalar field
quantized in the background of the $d-2$ - brane. Thus, the
latter vacuum is somewhat similar to the medium of classical
matter.

Although the strong and weak energy conditions are violated for all values of
$\xi$ in the case of massive scalar field quantized in the background of the
brane, the conformal coupling $(\xi=\xi_c)$ is distinctive in the massive case
also. The vacuum tensor components at $\xi=\xi_c$ are given by
Eqs.(\ref{e6})-(\ref{e8}). Qualitatively, the temporal component (energy
density) is positive power divergent at small distances from the brane,
decreases with the increase of the distance, passes zero, becomes negative and
reaches minimum at $r\sim m^{-1}$, then increases and reaches zero
asymptotically from below with exponential behaviour. This is distinct from the
conventional Casimir effect, which corresponds to the vacuum energy density
being space independent constant, either negative or positive \cite{Most}. We
present also expressions for the vacuum tensor components at half-integer
values of the brane flux in the case of physical values of space dimension
$(d=n \geq  2)$, see Eqs.(\ref{d4})-(\ref{d9}) and, generally,
Eqs.(\ref{ab6})-(\ref{ab11}).

It should be noted that temporal components of the vacuum tensor
 for the quantized scalar (with $\xi=\frac14$) and spinor matter in
 magnetic backgrounds in low-dimensional $(d=2,\, 3)$ spaces were
 considered in Refs.\cite{Cangemi, Fry, Dunne, BordagKr,
Scandurra} .
  Since the authors of these works are concerned with
 the case when the region of nonvanishing background field is of
 nonvanishing transverse size
  and is overlapped with the region of nonvanishing quantized
 matter, their results differ considerably from ours: in
 particular, the dependence on the flux of the background magnetic
  field is not periodic. Moreover, in their case in addition to
 the  vacuum energy of quantized matter also the classical energy
 of  magnetic background has to be consistently taken into
 account,  as it is done in Ref.\cite{BordagKr}. On the contrary,
 when the region of  nonvanishing background field is impenetrable
 for quantized matter, then the quantum and classical energies are
 stored in different non-overlapping parts of space. Certainly our
 neglect of  transverse size of the magnetic brane is an
 idealization which  allowed us to solve the problem analytically.
 But we believe that  this idealization, as in the case of the
 conventional  Bohm-Aharonov effect \cite{Aha}, grasps some essential
 features of  the more realistic case of the magnetic brane with
 finite transverse size and the quantized scalar field satisfying
 the boundary condition on the edge of the brane. In particular, it
 is  almost evident that then the vacuum energy-momentum tensor
  components will depend periodically on the brane flux and
 possess  large-distance asymptotics which is exponential for
 $m>0$, see  Eqs.(\ref{f1})-(\ref{f3}), and
 negative-power-behaved for $m=0$, see Eqs.(\ref{f4})-(\ref{f6})
 with limit $m\rightarrow 0$ taken first before
 $r\rightarrow\infty$. A less evident, but still rather plausible
 conjecture is that tensor  components will behave less
 divergent, as compared to $(r-r_B)^{-d-3}$ for the transverse
 angular and $(r-r_B)^{-d-1}$ for all other components, near brane
 edge $r=r_B$. And a  really challenging task is to find out
 whether the vacuum energy density integrated over transverse
 coordinates will appear to be finite and somehow capable to
 compensate the classical energy per transverse section of the
 brane.

  An intriguing question is about the underlying physics of
 the  negativeness of the vacuum energy density and the violation
 of  strong and weak energy conditions. Recent examples of
 low-dimensional  (without magnetic background) models \cite{Olum}
 indicate that the  negativeness of the vacuum energy density at
 large distances is  related to effectively perfect reflection of
 quantized matter at  small distances; the negativeness might
 persist self-consistently  for the whole, classical plus quantum,
 energy density. In this respect Ref.\cite{Sitenko21} is worth mentioning,
  where the  vacuum energy density for the
 quantized spinor matter in the same background, as in the present
 paper, but exclusively in the $d=2$  case, was considered. As is
 known, spinor field cannot be made › zero at the location of a
 singular magnetic brane (which is a  point in the $d=2$ case).
 The whole set of permissible boundary conditions is parametrized
 by real quantity $\Theta$, and at  $\cos \Theta<0$ a bound state
 appears in the gap between positive  and negative frequency
 continua \cite{Gerbert}. Thus, for sure, perfect reflection at
 the point of singularity is excluded at $\cos \Theta<0$. As is
 shown in Ref.\cite{Sitenko21}, namely at these values of $\Theta$
 the vacuum energy density is strictly positive at all  distances,
 whereas, otherwise, its behaviour is similar to that of  the
 vacuum energy density for the quantized scalar matter with
 $\xi=\xi_c$, see, qualitatively, solid curves on Fig.2.

   At last,
 it should be noted that pitfalls of the zeta function
 regularization procedure (see, e.g., Refs.\cite{Sitenko23, Farhi,
BordagGo})
  do not appear in the case of singular backgrounds
 \cite{Sitenko21, Bab}. Owing  to this circumstance, it is
 possible to show that the renormalized vacuum energy-momentum
 tensor in the present paper is independent of the choice of a
 regularization procedure. In fact, this has been already shown
 for its temporal component at $\xi=\frac14$ in Ref.\cite{Bab}.  In
 a similar way this can be done for other values of $\xi$, and for
 other components. The key point of
  Ref.\cite{Bab} is that the heat kernel in the presence of the  brane
 coincides actually with the heat kernel in its  absence, and
 then, basing on this fact, it can be proved that the  use of zeta
 function regularization leads to renormalized  quantities which
 are regularization procedure independent.

\section*{Acknowledgements}
We are grateful to Professor A.G. Sitenko for invaluable help and enlightening
discussions. The research was supported by INTAS (grant INTAS OPEN 00-00055)
and Swiss National Science Foundation (grant SCOPES 2000-2003 7 IP 62607).


{ \setcounter{equation}{0}
\renewcommand{\theequation}{A.\arabic{equation}}
\appendix
\section*{Appendix A}
Using Eq.(\ref{b38}), we present Eqs.(\ref{b28}) and (\ref{b29})
in the following form
\begin{multline}\label{ap1}
\Omega_0(s)=\frac{r\sin\left[\left(s-\frac
d2\right)\pi\right]}{\left(s-\frac d2\right)\pi}
\int\limits_0^\infty dy\,y^{d-2s+1}\int\limits_0^\infty
\frac{dk}{k^2+y^2+m^2}\times\\ \times
\left[J_F(kr)J_{-1+F}(kr)+J_{1-F}(kr)J_{-F}(kr)\right]\,,
\end{multline}
\begin{multline}\label{ap2}
{\tilde\Omega}_1(s)=\frac{2\sin\left[\left(1+s-\frac
d2\right)\pi\right]}{\pi} \int\limits_0^\infty
dy\,y^{d-2s-1}\int\limits_0^\infty \frac{dk\,k}{k^2+y^2+m^2}
\times\\ \times
\left\{\frac12(1+F)\left[J^2_{-F}(kr)-J^2_F(kr)\right]+\frac12(2-F)
\left[J^2_{-1+F}(kr)-J^2_{1-F}(kr)\right]-\right.\\
-\left.\frac{F(1-F)}{kr}\left[J_F(kr)J_{-1+F}(kr)+J_{1-F}(kr)J_{-F}(kr)\right]\right\}\,,
\end{multline}
\begin{multline}\label{ap3}
{\tilde\Omega}_2(s)=\frac{2\sin\left[\left(1+s-\frac
d2\right)\pi\right]}{\pi} \int\limits_0^\infty
dy\,y^{d-2s-1}\int\limits_0^\infty \frac{dk\,k}{k^2+y^2+m^2}
\times\\ \times
\left\{\frac12(1-F)\left[J^2_{-F}(kr)-J^2_F(kr)\right]+\frac12 F
\left[J^2_{-1+F}(kr)-J^2_{1-F}(kr)\right]+\right.\\
+\left.\frac{F(1-F)}{kr}\left[J_F(kr)J_{-1+F}(kr)+J_{1-F}(kr)J_{-F}(kr)\right]\right\}\,.
\end{multline}
Using relation (see Ref. \cite{Prud})
\begin{eqnarray}\label{ap4}
 \int\limits_0^\infty dk\,\frac{k^{\nu-\mu+1}}{k^2+\omega^2}J_\mu(kr)J_\nu(kr)=
 \omega^{\nu-\mu}I_\mu(\omega r)K_\nu(\omega r)\,,\quad
 -1<\nu<\mu+1\,,
\end{eqnarray}
 we get
\begin{multline}\label{ap5}
\Omega_0(s)=\frac{r\sin\left[\left(s-\frac
d2\right)\pi\right]}{\left(s-\frac d2\right)\pi}
 \int\limits_0^\infty dy
  \frac{y^{d-2s+1}}{\sqrt{m^2+y^2}} \times \\
  \times\left[I_F(r\sqrt{m^2+y^2})K_{1-F}(r\sqrt{m^2+y^2})+
  I_{1-F}(r\sqrt{m^2+y^2})K_F(r\sqrt{m^2+y^2})\right]\,.
\end{multline}
Introducing integration variable $w=r\sqrt{m^2+y^2}$ and using
identity
$$
 I_F(w)K_{1-F}(w)+I_{1-F}(w)K_F(w)=\frac 1w
 -\frac{2\sin(F\pi)}\pi K_F(w)K_{1-F}(w)\,,
$$
we get
\begin{multline}\label{ap6}
 \Omega_0(s)=\frac{m^{d-2s}}{2s-d}-
 \frac{4\sin(s-\frac d2) \pi}{(2s-d)\pi^2}\sin (F\pi)
 \,r^{2s-d}\times\\
 \times\int\limits_{mr}^\infty dw
 \left(w^2-m^2r^2\right)^{\frac d2 -s}
 K_F(w)K_{1-F}(w)\,.
\end{multline}

Although our derivation is valid for a strip $\frac d2<$
Re\,$s<\frac d2+1$ (see Eq.(\ref{b38})), the result,
Eq.(\ref{ap6}), is analytically continued to half-plane
Re\,$s<\frac d2+1$. Accordingly, $\Omega_0(s+1)$ is defined on
half-plane Re\,$s < \frac d2$. Taking relations
\begin{multline}\label{ap7}
t^{00}_{reg}(s)=\frac2{(4\pi)^{\frac
d2}}\frac{\Gamma\left(1+s-\frac d2\right)}{\Gamma(s)} \left[
\Omega_0(s)+\left(\frac14-\xi\right)\frac{1+s-\frac d2}s
\triangle_r  \Omega_0(s+1)\right]\,,
\end{multline}
\begin{multline}\label{ap8}
t^{jj}_{reg}(s)=\frac1{(4\pi)^{\frac
d2}}\frac{\Gamma\left(1+s-\frac d2\right)}{\Gamma(1+s)} \left[
\Omega_0(s)-\left(\frac14-\xi\right)2\left(1+s-\frac d2\right)
\triangle_r \Omega_0(s+1)\right]\,,
\end{multline}
into account, we obtain relations (\ref{b39}) and (\ref{b40}).

In a similar way as Eq.(\ref{ap6}), we get relations
\begin{multline}\label{ap9}
 {\tilde\Omega}_1(s)=-r^{-2}m^{d-2s-2}F(1-F)-\\
-\frac{2\sin(s-\frac d2) \pi}{\pi^2}\sin (F\pi)
 \,r^{2s-d}
\int\limits_{mr}^\infty dw \left(w^2-m^2r^2\right)^{\frac d2
-s-1}\times \\ \times\left[
 2F(1-F)K_F(w)K_{1-F}(w)+(1+F)w K_F^2(w)+
 (2-F)w K^2_{1-F}(w)\right]\,,
\end{multline}
\begin{multline}\label{ap10}
 {\tilde\Omega}_2(s)=r^{-2}m^{d-2s-2}F(1-F)+\\
+\frac{2\sin(s-\frac d2) \pi}{\pi^2}\sin (F\pi)
 \,r^{2s-d}
\int\limits_{mr}^\infty dw \left(w^2-m^2r^2\right)^{\frac d2
-s-1}\times \\ \times\left[
 2F(1-F)K_F(w)K_{1-F}(w)-(1-F)w K_F^2(w)-
 Fw K^2_{1-F}(w)\right]\,,
\end{multline}
which are analytically continued from strip $\frac d2-1<$
Re\,$s<\frac d2$ to half-plane Re\,$s<\frac d2$\,. Accordingly,
${\tilde\Omega}_1(s+1)$ and ${\tilde\Omega}_2(s+1)$ are defined
on half-plane Re\,$s<\frac d2-1$\,. Thus, we can obtain expression
for differences ${\tilde\Omega}_1(s)-m^2 {\tilde\Omega}_1(s+1)$
and ${\tilde\Omega}_2(s)-m^2 {\tilde\Omega}_2(s+1)$ which are
valid on half-plane Re\,$s<\frac d2-1$\,; note also that
contribution of first terms
 in the right-hand sides of Eqs.(\ref{ap9}) and (\ref{ap10}) is
 cancelled.

 Taking relations
 (\ref{b18}) and (\ref{b19}) into account, we obtain relations (\ref{b41}) and
 (\ref{b42}).
}


{ \setcounter{equation}{0}
\renewcommand{\theequation}{B.\arabic{equation}}
\section*{Appendix B}
Using repeatedly the recurrency relations for the incomplete
gamma functions, one can get
\begin{equation}\label{ab1}
\Gamma\left(-N-\frac12,
w\right)=\frac{(-1)^N}{\Gamma\left(N+\frac32\right)} \left[-\pi
\,\mathrm{erfc}\left(\sqrt{w}\,\right)+e^{-w}\sum_{l=0}^N (-1)^l
\Gamma\left(l+\frac12\right)w^{-l-\frac12}\right]\,,
\end{equation}
\begin{equation}\label{ab2}
\Gamma(-N-1, w)=\frac{(-1)^N}{\Gamma(N+2)}
\left[-E_1(w)+e^{-w}\sum_{l=0}^N (-1)^l
\Gamma(l+1)w^{-l-1}\right]\,,
\end{equation}
where
$$
\mathrm{erfc}(u)=\frac2{\sqrt{\pi}}\int\limits_u^\infty
d\tau\,e^{-\tau^2}
$$
is the complementary error function, and
$$
E_1(u)=\int\limits_u^\infty \frac{d\tau}\tau e^{-\tau}
$$
is the integral exponential function. Using Eqs.(\ref{ab1}), (\ref{ab2}) and
relations \cite{Prud}
\begin{multline}\label{ab3}
\int\limits_0^\infty d\tau\,
\tau^{-\frac12}e^{-\tau}\,\mathrm{erfc}\left(\frac{mr}{\sqrt{\tau}}\right)=\\
=\sqrt{\pi}\left\{1-2mr\left[K_0(2mr)L_{-1}(2mr)+K_1(2mr)L_0(2mr)\right]\right\},
\end{multline}
\begin{equation}\label{ab4}
\int\limits_0^\infty d\tau\,
\tau^{-\frac12}e^{-\tau}E_1\left(\frac{m^2r^2}\tau\right)=
2\sqrt{\pi}E_1(2mr)\,,
\end{equation}
we perform integration in Eqs.(\ref{d1}) - (\ref{d3}). Further, in the case of
odd space dimension we use representation of the Macdonald function of
half-integer order through a finite sum:
\begin{equation}\label{ab5}
K_{l+\frac12}(u)=\sqrt{\pi}e^{-u}\sum_{n=0}^l
\frac{\Gamma(l+n+1)}{\Gamma(n+1)\Gamma(l-n+1)}(2u)^{-n-\frac12}\,.
\end{equation}

Consequently, we get
\begin{multline}\label{ab6}
\left.t^{00}_{ren}\right|_{F=\frac12}=
\frac{2m^{2N+1}}{(4\pi)^{N+\frac12}}
\frac{(-1)^N}{\Gamma\left(N+\frac32\right)} \left\{ -\frac\pi2
+\pi mr\left[K_0(2mr)L_{-1}(2mr)+\right.\right.\\
\left.+K_1(2mr)L_0(2mr)\right]
+\frac1{\sqrt{\pi}}\sum_{l=0}^{N-2}(-1)^l\Gamma\left(l+\frac12\right)(mr)^{-l}K_{l+1}(2mr)-\\
-\frac{(-1)^N}{\sqrt{\pi}}\Gamma\left(N-\frac12\right)
\left[1-\frac14(1-4\xi)(4N^2-1)(mr)^{-2}\right](mr)^{-N+1}K_N(2mr)+\\
\left.+\frac{(-1)^N}{\sqrt{\pi}}\Gamma\left(N+\frac12\right)
\left[1-(1-4\xi)(2N+1)\right](mr)^{-N}K_{N+1}(2mr)
\right\}\,,\quad d=2N\,,
\end{multline}
\begin{multline}\label{ab7}
\left.t^{rr}_{ren}\right|_{F=\frac12}=
-\frac{2m^{2N+1}}{(4\pi)^{N+\frac12}}
\frac{(-1)^N}{\Gamma\left(N+\frac32\right)} \left\{ -\frac\pi2
+\pi mr\left[K_0(2mr)L_{-1}(2mr)+\right.\right.\\
\left.+K_1(2mr)L_0(2mr)\right]
+\frac1{\sqrt{\pi}}\sum_{l=0}^{N-2}(-1)^l\Gamma\left(l+\frac12\right)(mr)^{-l}K_{l+1}(2mr)-\\
-\frac{(-1)^N}{\sqrt{\pi}}\Gamma\left(N-\frac12\right)
\left[1+\xi(4N^2-1)(mr)^{-2}\right](mr)^{-N+1}K_N(2mr)+\\
\left.+\frac{(-1)^N}{\sqrt{\pi}}\Gamma\left(N+\frac12\right)
(mr)^{-N}K_{N+1}(2mr) \right\}\,,\quad d=2N\,,
\end{multline}
\begin{multline}\label{ab8}
\left.t^{\varphi\varphi}_{ren}\right|_{F=\frac12}=
-\frac{2m^{2N+1}}{r^2(4\pi)^{N+\frac12}}
\frac{(-1)^N}{\Gamma\left(N+\frac32\right)} \left\{ -\frac\pi2 +
\pi mr\left[K_0(2mr)L_{-1}(2mr)+\right.\right.\\
\left.+K_1(2mr)L_0(2mr)\right]
+\frac1{\sqrt{\pi}}\sum_{l=0}^{N-1}(-1)^l\Gamma\left(l+\frac12\right)(mr)^{-l}K_{l+1}(2mr)+\\
\left.+\frac{(-1)^N}{\sqrt{\pi}}\Gamma\left(N+\frac12\right)
\left[1-(1-4\xi)(2N+1)\right](mr)^{-N}K_{N+1}(2mr)
\right\}\,,\quad d=2N\,,
\end{multline}
\begin{multline}\label{ab9}
\left.t^{00}_{ren}\right|_{F=\frac12}=
\frac{2m^{2N+2}}{(4\pi)^{N+1}}
\left\{-\frac{(-1)^N}{\Gamma(N+2)}E_1(2mr)+\right.\\
+\frac{e^{-2mr}}{\Gamma(N+2)}\sum_{l=0}^{N-1}(-1)^{N-l}\Gamma(l+1)\sum_{n=0}^{l+1}
\frac{\Gamma(l+n+2)(mr)^{-l-n-1}}{2^{2n+1}\Gamma(n+1)\Gamma(l-n+2)}+\\
+(1-4\xi)e^{-2mr}\sum_{l=0}^N
\frac{\Gamma(N+l+1)(mr)^{-N-l-2}}{2^{2l+1}\Gamma(l+1)\Gamma(N-l+1)}+\\
\left.+\left[\frac1{N+1}-2(1-4\xi)\right]e^{-2mr}\sum_{l=0}^{N+1}
\frac{\Gamma(N+l+2)(mr)^{-N-l-1}}{2^{2l+1}\Gamma(l+1)\Gamma(N-l+2)}\right\}
\,,\quad d=2N+1\,,
\end{multline}
\begin{multline}\label{ab10}
\left.t^{rr}_{ren}\right|_{F=\frac12}=
-\frac{2m^{2N+2}}{(4\pi)^{N+1}}
\left\{-\frac{(-1)^N}{\Gamma(N+2)}E_1(2mr)+\right.\\
+\frac{e^{-2mr}}{\Gamma(N+2)}\sum_{l=0}^{N-1}(-1)^{N-l}\Gamma(l+1)\sum_{n=0}^{l+1}
\frac{\Gamma(l+n+2)(mr)^{-l-n-1}}{2^{2n+1}\Gamma(n+1)\Gamma(l-n+2)}-\\
-4\xi e^{-2mr}\sum_{l=0}^N
\frac{\Gamma(N+l+1)(mr)^{-N-l-2}}{2^{2l+1}\Gamma(l+1)\Gamma(N-l+1)}+\\
\left.+\frac{e^{-2mr}}{N+1}\sum_{l=0}^{N+1}
\frac{\Gamma(N+l+2)(mr)^{-N-l-1}}{2^{2l+1}\Gamma(l+1)\Gamma(N-l+2)}\right\}
\,,\quad d=2N+1\,,
\end{multline}
\begin{multline}\label{ab11}
\left.t^{\varphi\varphi}_{ren}\right|_{F=\frac12}=
-\frac{2m^{2N+2}}{r^2(4\pi)^{N+1}}
\left\{-\frac{(-1)^N}{\Gamma(N+2)}E_1(2mr)+\right.\\
+\frac{e^{-2mr}}{\Gamma(N+2)}\sum_{l=0}^{N-1}(-1)^{N-l}\Gamma(l+1)\sum_{n=0}^{l+1}
\frac{\Gamma(l+n+2)(mr)^{-l-n-1}}{2^{2n+1}\Gamma(n+1)\Gamma(l-n+2)}+\\
\left.+\left[\frac1{N+1}-2(1-4\xi)\right]e^{-2mr}\sum_{l=0}^{N+1}
\frac{\Gamma(N+l+2)(mr)^{-N-l-1}}{2^{2l+1}\Gamma(l+1)\Gamma(N-l+2)}\right\}
\,,\quad d=2N+1\,.
\end{multline}
 }



\begin{figure}
\begin{tabular}{c}
\includegraphics[width=140mm]{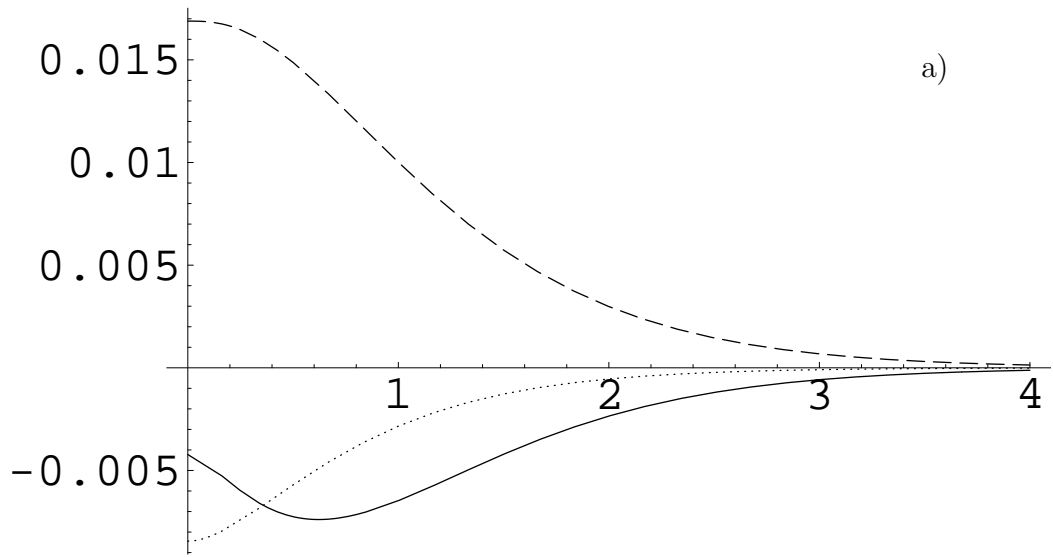}
\put(-50,200){a)}
\end{tabular}
\begin{tabular}{c}
\includegraphics[width=140mm]{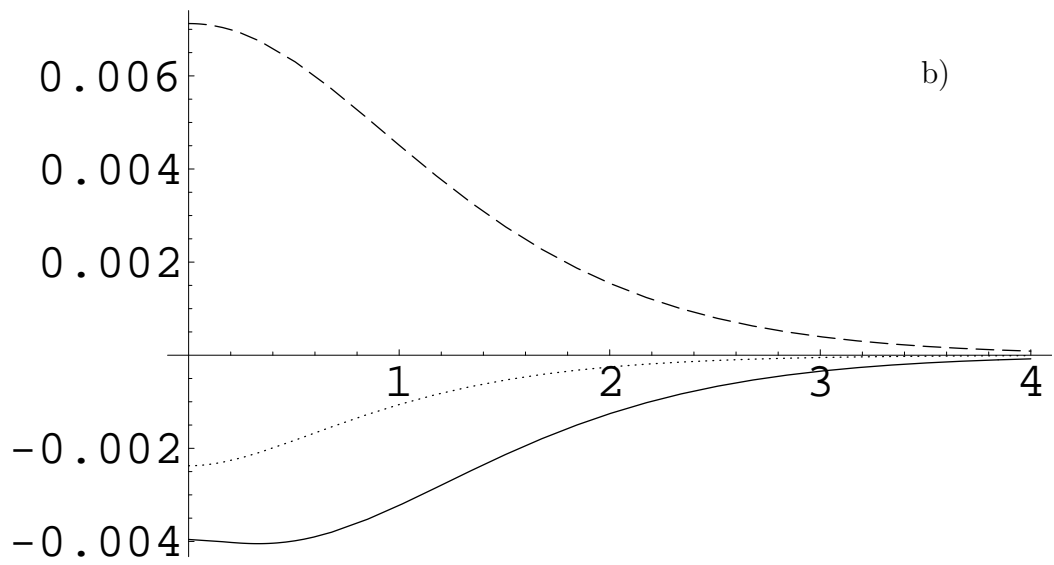}
\put(-50,200){b)}
\end{tabular}
\caption{$\xi=0$\quad a) $d=2$,\quad b) $d=3$.}
\end{figure}

\begin{figure}
\begin{tabular}{c}
\includegraphics[width=140mm]{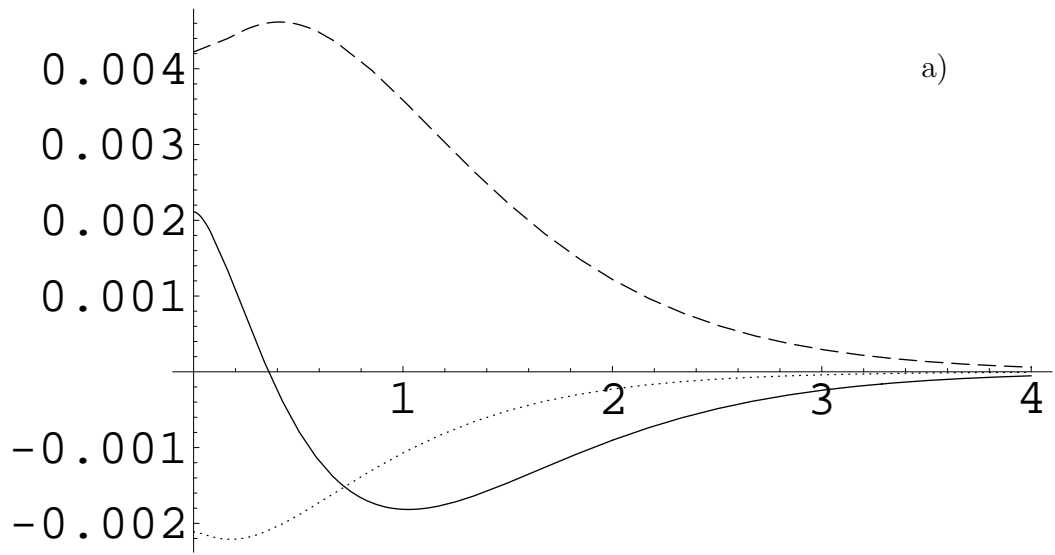}
\put(-50,200){a)}
\end{tabular}
\begin{tabular}{c}
\includegraphics[width=140mm]{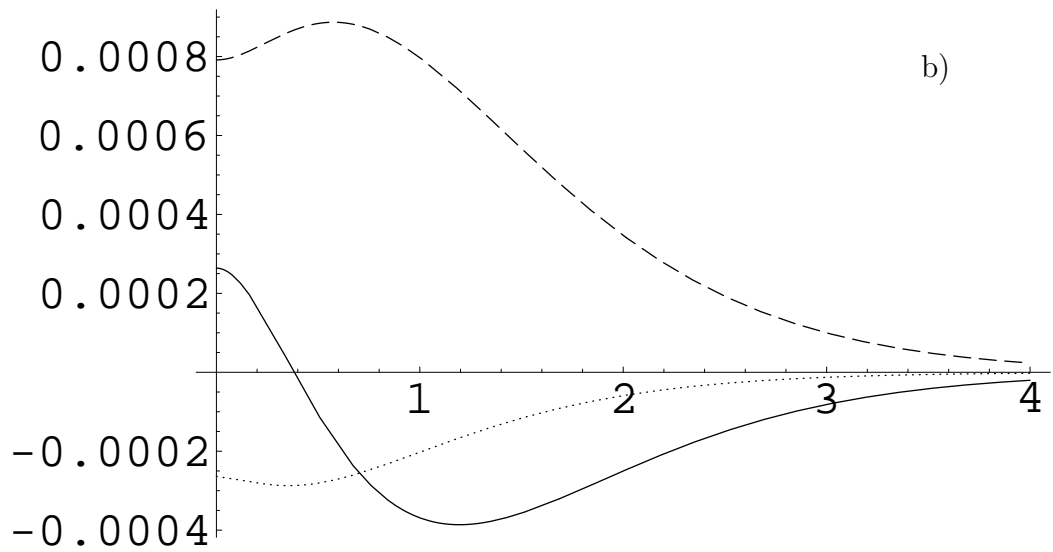}
\put(-50,200){b)}
\end{tabular}
\caption{$\xi=\xi_c$\quad a) $d=2$,\quad b) $d=3$.}
\end{figure}

\begin{figure}
\begin{tabular}{c}
\includegraphics[width=140mm]{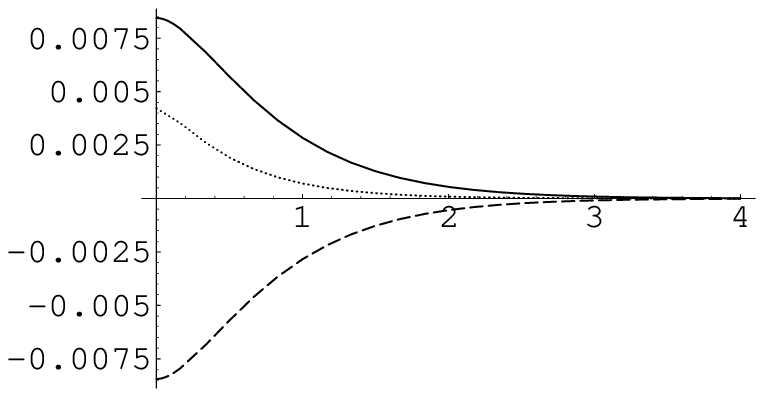}
\put(-50,200){a)}
\end{tabular}
\begin{tabular}{c}
\includegraphics[width=140mm]{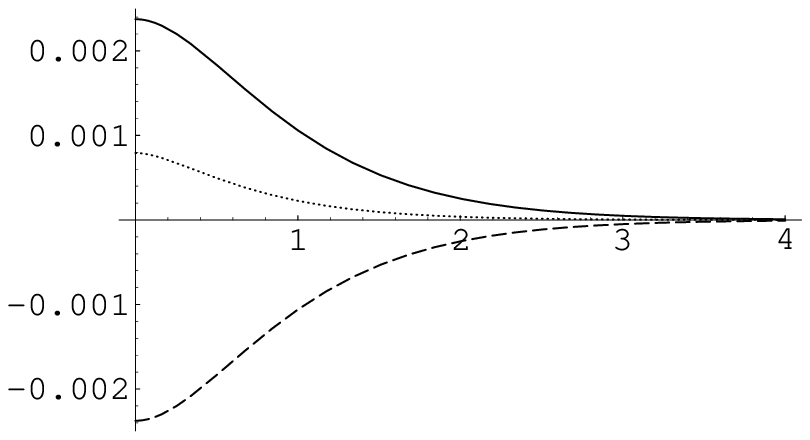}
\put(-50,200){b)}
\end{tabular}
\caption{$\xi=1/4$\quad a) $d=2$,\quad b) $d=3$.}
\end{figure}


\begin {thebibliography}{99}
\raggedright
\bibitem{Cas} H.B.G. Casimir, Proc. Kon. Ned. Akad. Wetenschap B \textbf{51},
793 (1948); Physica \textbf{19}, 846 (1953).
\bibitem{Bir} N.D. Birrel and P.C.W. Davies, \textit{Quantum Fields in
Curved Space}  (Cambridge Univ. Press, Cambridge, 1982).
\bibitem{Most} V.M. Mostepanenko and N.N. Trunov, \textit{The Casimir Effect and Its Applications}
(Clarendon Press, Oxford, 1997).
\bibitem{Star}A.A. Starobinsky, Phys. Lett. B \textbf{91}, 99 (1980); Ya.B.
Zeldovich and A.A. Starobinsky, Astron. Lett. \textbf{10}, 135
(1984).
\bibitem{Aha} Y. Aharonov and D. Bohm, Phys. Rev. \textbf{115}, 485 (1959).
\bibitem{Sit} Yu.A. Sitenko and A.Yu. Babansky, Mod. Phys. Lett. A
\textbf{13}, 379 (1998); Phys. Atom. Nucl. \textbf{61}, 1594
(1998).
\bibitem{Pen} R. Penrose, in: \textit{Relativity, Groups and
Topology}, edited by B.S. DeWitt, C. DeWitt (Gordon and Breach,
New York, 1964).
\bibitem{Cher} N.A. Chernikov and E.A. Tagirov, Ann. Inst. Henri
Poincare \textbf{9} A, 109 (1968).
\bibitem{Cal} C.G. Callan, S. Coleman, and R. Jackiw, Ann. Phys. (N.Y.)
\textbf{59}, 42 (1970).
\bibitem{Sal} A. Salam and J. Strathdee, Nucl. Phys. B \textbf{90}, 203
(1975).
\bibitem{Dow} J.S. Dowker and R. Critchley, Phys. Rev. D \textbf{13}, 3224
(1976).
\bibitem{Haw} S.W. Hawking, Commun. Math. Phys. \textbf{55}, 133 (1977).
\bibitem{Prud} A.P. Prudnikov, Yu.A. Brychkov, and O.I.
Marychev, \textit{Integrals and Series: Special Functions} (Gordon
and Breach, New York, 1986).
\bibitem{Abra} \textit{Handbook of Mathematical Functions},
 edited by M. Abramowitz and I.A. Stegun (Dover, New York, 1972).
\bibitem{HaE} S.W. Hawking and G.F.R. Ellis, \textit{The Large
Scale Structure of Space-Time} (Cambridge Univ. Press, Cambridge,
1973).
\bibitem{Cangemi} D. Cangemi, G. Dunne, and E. D'Hoker, Phys. Rev. D \textbf{52}, 3163 (1995).
\bibitem{Fry} M.P. Fry, Phys. Rev. D \textbf{54}, 6444 (1996).
\bibitem{Dunne}  G. Dunne and T.M. Hall, Phys. Lett. B \textbf{419}, 322 (1998).
\bibitem{BordagKr} M. Bordag and K. Kirsten, Phys. Rev. D \textbf{60}, 105019 (1999).
\bibitem{Scandurra} M. Scandurra, Phys. Rev. D \textbf{62}, 085024 (2000).
\bibitem{Olum} K.D. Olum and N. Graham,
         \textit{Static Negative Energies near a Domain Wall}, e-Print
         Archive: gr-qc/0205134;

         N. Graham and K.D. Olum,
         \textit{Negative Energy Densities in Quantum Field Theory with a
          Background Potential}, e-Print Archive: hep-th/0211244.
\bibitem{Sitenko21} Yu.A. Sitenko, Phys.Atom.Nucl. \textbf{62}, 1056 (1999).
\bibitem{Gerbert}  P. de Sousa Gerbert, Phys. Rev. D \textbf{40}, 1346 (1989).
\bibitem{Sitenko23} Yu.A. Sitenko and D.G. Rakityansky, Phys.Atom.Nucl.  \textbf{61}, 790 (1998).
\bibitem{Farhi} E. Farhi, N. Graham, R.L. Jaffe, and H. Weigel,
                Nucl. Phys. B \textbf{595}, 536 (2001).
\bibitem{BordagGo} M. Bordag, A.S. Goldhaber, P. van Nieuwenhuizen, and D.Vassilevich,
                Phys. Rev. D \textbf{66}, 125014  (2002).
\bibitem{Bab} A.Yu. Babansky and Yu.A. Sitenko, Theor. Math.
              Phys. \textbf{120}, 876 (1999).
\end{thebibliography}
\end{document}